\documentclass[sigconf]{acmart}

\AtBeginDocument{%
  }

\setcopyright{acmlicensed}
\copyrightyear{2018}
\acmYear{2018}
\acmDOI{XXXXXXX.XXXXXXX}
\acmConference[Conference acronym 'XX]{Make sure to enter the correct
  conference title from your rights confirmation email}{June 03--05,
  2018}{Woodstock, NY}
\acmISBN{978-1-4503-XXXX-X/2018/06}

\usepackage{listings}
\usepackage{caption}

\usepackage{xcolor,colortbl}
\definecolor{pshigh}{rgb}{1.0,0.5,0.5}
\definecolor{phigh}{rgb}{1.0,0.6,0.6}
\definecolor{pmiddle}{rgb}{1.0,0.8,0.8}
\definecolor{plow}{rgb}{1.0,0.85,0.85}
\definecolor{pslow}{rgb}{1.0,0.9,0.9}
\definecolor{nshigh}{rgb}{0.7,0.7,1.0}
\definecolor{nhigh}{rgb}{0.75,0.75,1.0}
\definecolor{nmiddle}{rgb}{0.8,0.8,1.0}
\definecolor{nlow}{rgb}{0.85,0.85,1.0}
\definecolor{nslow}{rgb}{0.9,0.9,1.0}
\begin{document}

\title{Revisiting Prompt Engineering: A Comprehensive Evaluation for LLM-based Personalized Recommendation}

\author{Genki Kusano}
\email{g-kusano@nec.com}
\affiliation{%
  \institution{NEC Corporation}
  \state{Kawasaki}
  \country{Japan}
}

\author{Kosuke Akimoto}
\email{kosuke\_a@nec.com}
\affiliation{%
  \institution{NEC Corporation}
  \state{Kawasaki}
  \country{Japan}
}

\author{Kunihiro Takeoka}
\email{k\_takeoka@nec.com}
\affiliation{%
  \institution{NEC Corporation}
  \state{Kawasaki}
  \country{Japan}
}

\renewcommand{\shortauthors}{Kusano et al.}
\renewcommand{\shorttitle}{Revisiting Prompt Engineering for Personalized Recommendation}

\begin{abstract}
Large language models (LLMs) can perform recommendation tasks by taking prompts written in natural language as input. Compared to traditional methods such as collaborative filtering, LLM-based recommendation offers advantages in handling cold-start, cross-domain, and zero-shot scenarios, as well as supporting flexible input formats and generating explanations of user behavior. In this paper, we focus on a single-user setting, where no information from other users is used. This setting is practical for privacy-sensitive or data-limited applications. In such cases, prompt engineering becomes especially important for controlling the output generated by the LLM. We conduct a large-scale comparison of 23 prompt types across 8 public datasets and 12 LLMs. We use statistical tests and linear mixed-effects models to evaluate both accuracy and inference cost. Our results show that for cost-efficient LLMs, three types of prompts are especially effective: those that rephrase instructions, consider background knowledge, and make the reasoning process easier to follow. For high-performance LLMs, simple prompts often outperform more complex ones while reducing cost. In contrast, commonly used prompting styles in natural language processing, such as step-by-step reasoning, or the use of reasoning models often lead to lower accuracy. Based on these findings, we provide practical suggestions for selecting prompts and LLMs depending on the required balance between accuracy and cost.

\textcolor{red}{This paper has been accepted to ACM RecSys 2025. Please cite it appropriately after September 22, 2025.}
\end{abstract}


\begin{CCSXML}
<ccs2012>
   <concept>
       <concept_id>10002951.10003317.10003347.10003350</concept_id>
       <concept_desc>Information systems~Recommender systems</concept_desc>
       <concept_significance>500</concept_significance>
       </concept>
 </ccs2012>
\end{CCSXML}

\ccsdesc[500]{Information systems~Recommender systems}

\keywords{LLM-based Recommendation, Prompt Evaluation, Statistical Analysis, Reasoning}

\maketitle

\section{Introduction}

% Why LLM-based recommendation matters
Recommender systems are widely used in services such as online advertising, e-commerce, and video streaming. Traditional approaches, such as collaborative filtering, predict user behavior by learning patterns from many users' histories~\cite{DBLP:conf/recsys/CovingtonAS16,DBLP:journals/computer/KorenBV09,DBLP:conf/www/HeLZNHC17_ncf}. 
These methods work well when there is a large amount of training data. However, their performance decreases in cold-start situations~\cite{DBLP:conf/sigir/ZhuSSC20_heater,DBLP:conf/mm/WeiWLNLLC21_clcrec,DBLP:conf/www/ZhouZY23_ccfcrec} where the system has not seen the target items or users before, and also in cross-domain situations~\cite{DBLP:journals/tois/ZangZLZY23_cross_domain_survery,DBLP:journals/csur/KhanIG17_cdr_survey,DBLP:conf/ijcai/ManSJC17_emcdr} where the system is applied to a different domain from its original training domain, such as recommending books after learning from movie data.

\begin{figure}[ht]
\centering
\includegraphics[width=85mm]{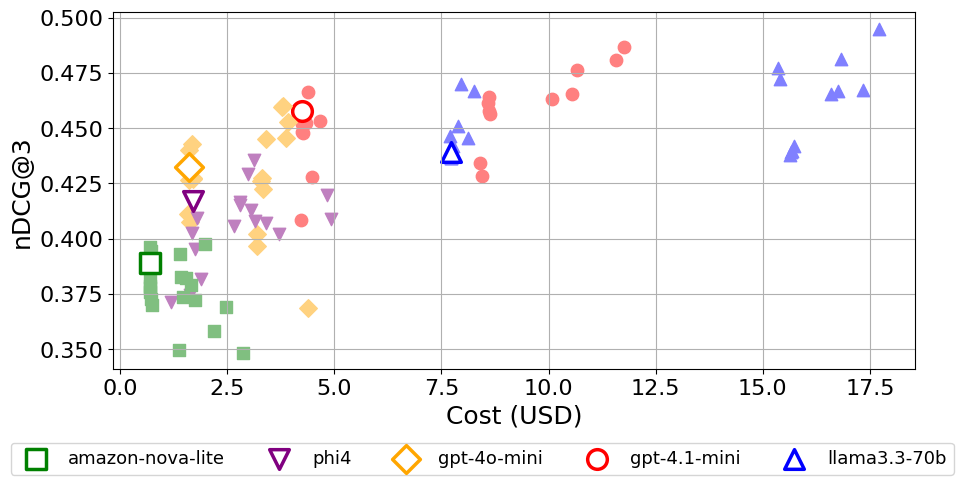}
\caption{Recommendation accuracy (nDCG@$3$) and inference cost (USD for processing 1,600 users from 8 datasets) for 21 prompts, including the baseline, evaluated with five LLMs in Section~\ref{sec:prompt-level}. Hollow and filled markers represent the baseline and 20 prompts from previous studies, respectively.}
\label{fig:cost_map_middle}
\end{figure}

A new approach that can address these limitations is the use of large language models (LLMs). When a recommendation task is described in natural language, an LLM can return meaningful results without additional training~\cite{DBLP:conf/kdd/DeldjooHMKSRVSK24_survey,DBLP:journals/tkde/Fan24_survey,DBLP:conf/coling/LiZLC24_survey,DBLP:journals/tois/Jianghao24_survey,DBLP:journals/www/WuZQWGSQZZLXC24_survey,xu2025tappingpotentiallargelanguage,DBLP:journals/corr/cotrec-llm}. These models support a wide range of input formats and generate fluent responses, which was not possible in earlier methods. They can also handle item attributes that are not included in the training data and provide informative explanations for their inferences.

% Prompt design issues for LLM recommendation
Prompt engineering is also used in recommendation, although it does not always lead to better accuracy. For example, techniques developed in natural language processing (NLP), such as step-by-step reasoning~\cite{DBLP:conf/nips/KojimaGRMI22_zscot} and rephrasing~\cite{DBLP:journals/corr/abs-2311-04205_rephrase}, have been effective in tasks like question answering and summarization. Recommendation tasks, however, require different types of reasoning that emphasize the relationship between users and items. Several studies have examined prompts in both general NLP~\cite{DBLP:journals/corr/abs-2407_12994_prompt_survey_0, DBLP:journals/corr/abs-2402-07927_prompt_survey_51} and recommendation~\cite{xu2025tappingpotentiallargelanguage,DBLP:journals/corr/23_is_chatgpt,DBLP:journals/corr/abs-2303-14524_chatrec,DBLP:conf/recsys/DaiSZYSXS0X23_listwise,DBLP:journals/corr/23_summary,DBLP:conf/www/ZhengCQZ024_block_summary,DBLP:conf/naacl/WangL24_demo,DBLP:conf/ecir/HouZLLXMZ24_sort,DBLP:conf/cikm/HeXJSLFMKM23_zeroshotconversational}. Research on recommendation has mostly focused on prompt types proposed in the RecSys field. It often includes few techniques from NLP and covers a limited range of datasets or LLMs. As a result, the conclusions from these studies may not generalize well to broader recommendation settings.

% Why focus on single-user (personalized) setting
To address these gaps, we conduct a broad evaluation of prompts for LLM-based recommendation. Our study compares \textbf{23 prompt types, 8 real-world datasets, and 12 LLMs}, which is much larger in scale than previous studies. This paper focuses on a personalized setting, where the LLM uses only one user's interaction history and does not rely on data from others. While incorporating other users may add information, it would also increase the complexity of the experimental design. By limiting the input to a single user, we can more directly assess the effect of each prompt on recommendation performance, without being influenced by other users. 

% Evaluation and insights
In the experiments, we first evaluated five cost-efficient LLMs (Figure~\ref{fig:cost_map_middle}), such as \texttt{gpt-4.1-mini} and \texttt{llama3.3-70b}. We used hypothesis testing and linear mixed-effects models to identify which prompts led to better recommendation accuracy. The results showed that three prompt types worked well across different datasets and LLMs. These prompts encouraged rephrasing the input~\cite{DBLP:journals/corr/abs-2311-04205_rephrase}, considering background knowledge~\cite{DBLP:conf/iclr/ZhengMCCCLZ24_stepback}, or clarifying the task goal using a specific thinking process~\cite{DBLP:conf/iclr/YaoZYDSN023_react}. In contrast, step-by-step reasoning~\cite{DBLP:conf/nips/KojimaGRMI22_zscot}, which is often said to be effective in NLP, did not lead to higher accuracy in recommendation.

We then analyzed which combinations of prompts and LLMs were effective in terms of both accuracy and inference cost. In this experiment, we added high-performance LLMs (e.g., \texttt{gpt-4.1} and \texttt{claude-3.7-sonnet}) and reasoning models (e.g., \texttt{o3} and \texttt{o4-mini}) to the comparison. In accuracy-focused cases, \texttt{claude-3.7-sonnet} reached the highest accuracy. However, the cost increased sharply when more complex prompts were used, even though their accuracy was similar to that of the simple baseline prompt. For this reason, \textbf{if accuracy is the top priority, using the simple baseline prompt with \texttt{claude-3.7-sonnet} is more effective than relying on more complex prompt designs.} On the other hand, \textbf{if cost is also a concern, a better option is to use \texttt{gpt-4.1-mini} or \texttt{llama3.3-70b} with one of the three effective prompts}~\cite{DBLP:journals/corr/abs-2311-04205_rephrase, DBLP:conf/iclr/ZhengMCCCLZ24_stepback, DBLP:conf/iclr/YaoZYDSN023_react}. This combination reached about 90\% of the accuracy of \texttt{claude-3.7-sonnet}, at less than one fifth of the cost.

As an additional observation, we found that reasoning models such as \texttt{o3} showed relatively strong performance. However, their accuracy did not exceed that of \texttt{claude-3.7-sonnet}, and the inference cost was higher. To explore the impact of reasoning, we further evaluated internal reasoning settings by activating the thinking mode in \texttt{claude-3.7-sonnet} and applying the three effective prompts to \texttt{o3-mini}. In some cases, these settings led to lower accuracy than the baseline prompt. These results suggest that \textbf{adding more reasoning does not always improve performance in recommendation tasks and may sometimes reduce accuracy}.

% Contributions and conclusion
This study provides practical guidance on selecting prompts and LLMs for personalized LLM-based recommendation. All code is released on \url{https://github.com/nec-research-labs/recsys2025_reproducibility_prompts}.

\section{Related Work}
{\em NLP-Based Recommendation.} Natural language processing (NLP) has contributed to the development of recommender systems. Earlier techniques applied word embeddings such as word2vec to represent items based on user interactions, as seen in Prod2Vec~\cite{DBLP:conf/kdd/GrbovicRDBSBS15_prod2vec} and Item2Vec~\cite{DBLP:conf/recsys/BarkanK16_item2vec}. Subsequent approaches introduced transformer-based models such as BERT to handle items as tokens, as in BERT4Rec~\cite{DBLP:conf/cikm/SunLWPLOJ19_bert4rec}. LLMs have built on this work by providing contextual understanding of both users and items and by generating outputs in natural language. Pretrained language models like T5 or LLaMA have been used to encode user reviews and product descriptions~\cite{DBLP:journals/corr/23_llamarec, DBLP:conf/recsys/BaoZZWF023_tallrec, DBLP:conf/cikm/LiZC23_p5}.

{\em Prompt Engineering for Recommendation.} Prompt-based recommendation has received attention for its ability to perform without labeled data or fine-tuning, making it suitable for zero-shot settings. Early studies applied generative models such as ChatGPT to recommendation tasks and explored various prompt styles~\cite{DBLP:journals/corr/23_is_chatgpt, DBLP:journals/corr/abs-2303-14524_chatrec}. Subsequent work proposed methods to improve accuracy, including listwise input formats~\cite{DBLP:conf/recsys/DaiSZYSXS0X23_listwise}, user history summarization~\cite{DBLP:journals/corr/23_summary,DBLP:conf/www/ZhengCQZ024_block_summary}, in-context learning with demonstrations~\cite{DBLP:conf/naacl/WangL24_demo}, emphasis on recent user behavior~\cite{DBLP:conf/ecir/HouZLLXMZ24_sort}, and dialogue-based modeling~\cite{DBLP:conf/cikm/HeXJSLFMKM23_zeroshotconversational}.

{\em Evaluation in Recommendation.} Reliable and reproducible evaluation is important for comparing approaches in recommendation. Previous studies have reexamined conventional methods such as matrix factorization~\cite{DBLP:conf/recsys/progress_DacremaCJ19,DBLP:conf/recsys/SunY00Q0G20_reproduce,DBLP:conf/recsys/RendleKZA20_ncf,DBLP:conf/recsys/revisit_MilogradskiiLPA24} and sequential recommendation~\cite{DBLP:journals/access/BetelloPSTBTS25_srs,DBLP:conf/recsys/PetrovM22a_srs_bert,DBLP:conf/recsys/DallmannZH21_srs_item}, focusing on their reproducibility. Several studies have examined LLM-based approaches~\cite{xu2025tappingpotentiallargelanguage,DBLP:journals/tkde/Fan24_survey,DBLP:journals/tois/Jianghao24_survey,DBLP:conf/kdd/DeldjooHMKSRVSK24_survey}, but they used only a small number of prompts or datasets. This paper presents a broader and more systematic comparison across different prompts, datasets, and LLMs.

\section{Experimental Settings}
\label{sec:experiment}
To examine how prompt design influences recommendation accuracy, we conducted large-scale experiments using eight datasets, two user types, 12 LLMs, and standard ranking metrics.

\subsection{Datasets}
\label{subsec:datasets}

We used eight datasets: the Yelp Dataset\footnote{\url{https://www.kaggle.com/datasets/yelp-dataset/yelp-dataset}}, the Microsoft News Recommendation Dataset (MIND)~\cite{DBLP:conf/acl/WuQCWQLLXGWZ20_mind}, the Food Dataset~\cite{DBLP:conf/emnlp/MajumderLNM19_food}, and five categories from the Amazon Review Dataset~\cite{DBLP:conf/emnlp/amazon_NiLM19}: Music, Movies, Groceries, Clothes, and Books.

In each dataset, users with 6 to 11 interactions were labeled as {\em light}, and those with 31 or more as {\em heavy}. The most recent item\footnote{We use the term {\em item} to refer to entities such as news in MIND, recipes in Food, or businesses in Yelp.} was used as the positive example for inference, and the remaining interactions were used to construct the user prompt. We randomly selected 100 users for each dataset and user type, resulting in 1,600 users in total (8 datasets $\times$ 2 types $\times$ 100 users).

To fairly compare prompt performance, we fixed the same ranking problem across all prompts by using a set of 10 items: one positive and nine negatives. Negative items were randomly selected from those the user had not interacted with. For MIND, these were news articles delivered but not clicked. For the Yelp dataset, we used businesses in the same state as the positive item. For the remaining datasets, we sampled items not included in the user's history.

\subsection{LLMs}

We evaluated 12 LLMs from OpenAI, Meta, AWS, Microsoft, and Anthropic by fixing the temperature at $0.1$. Based on cost and internal behavior, we classified the LLMs into three: low-cost, high-cost, and reasoning models. Table~\ref{table:llms} summarizes the LLMs, including their versions, cost classes, and token prices. An LLM is classified as high-cost if its output token price exceeds 2 US dollars. \texttt{claude-3.7-sonnet} is used as a standard LLM by default, but treated as a reasoning model when the ``thinking mode'' is enabled. Reasoning models typically involve additional cost due to internal multi-step inference, and their actual cost is often several times higher than the listed token price.

All OpenAI models were accessed through the official API. Models from Meta, AWS, and Anthropic were accessed via Amazon Bedrock. \texttt{phi-4} was run locally, and its cost was estimated based on publicly available API pricing\footnote{\url{https://deepinfra.com/microsoft/phi-4}. All pricing in Table~\ref{table:llms} is as of March 13, 2025.}.

\begin{table}[htbp]
\centering
\caption{LLMs used in this study, including their version, cost classes, and token prices (USD per 1M tokens).}
\label{table:llms}
\scalebox{0.74}{
\begin{tabular}{llllcc}
\toprule
\textbf{Provider} & \textbf{Model} & \textbf{Version} & \textbf{Cost Class} & \textbf{Input} & \textbf{Output} \\
\midrule
OpenAI     & \texttt{gpt-4o-mini}       & 2024-07-18      & Low        & 0.15 & 0.6  \\
OpenAI     & \texttt{gpt-4.1-mini}      & 2025-04-14      & Low        & 0.4  & 1.6  \\
OpenAI     & \texttt{gpt-4.1}           & 2025-04-14      & High       & 2.0  & 8.0  \\
OpenAI     & \texttt{o3-mini}           & 2025-01-31      & Reasoning  & 1.1  & 4.0  \\
OpenAI     & \texttt{o4-mini}           & 2025-04-16      & Reasoning  & 1.1  & 4.0  \\
OpenAI     & \texttt{o3}                & 2025-04-16      & Reasoning  & 10   & 40   \\
Meta       & \texttt{llama3.3-70b}      & v1(2024-12-19)              & Low        & 0.72 & 0.72 \\
AWS        & \texttt{amazon-nova-lite} & v1(2024-12-04)              & Low        & 0.06 & 0.24 \\
Microsoft  & \texttt{phi-4}             & 2024-12-12   & Low        & 0.07 & 0.14 \\
Anthropic  & \texttt{claude-3.5-haiku}  & 20241022-v1     & High       & 0.8  & 4.0  \\
Anthropic  & \texttt{claude-3.7-sonnet} & 20250219-v1     & High      & 3.0  & 15   \\
\bottomrule
\end{tabular}
}
\end{table}

\subsection{Prompts}
\label{subsec:prompt}

This section describes the prompts compared in this paper. LLMs typically use three main roles in conversation: \texttt{system}, \texttt{user}, and \texttt{assistant}, as seen in OpenAI and Anthropic models. Other LLMs, such as \texttt{llama3.3-70b} and \texttt{phi-4}, do not follow the same role names but support similar role-based prompting. For consistency, we use \texttt{system}, \texttt{user}, and \texttt{assistant} to describe all prompts in this paper. These LLMs also support multi-turn input, where earlier messages influence the response by providing context. Our experiments evaluate both single-turn and multi-turn prompt formats.

\subsubsection{Standardized Phrases}
To ensure fair evaluation across different prompts, we use a fixed wording pattern for all prompts, as shown in Table~\ref{table:prompt_components}. The variables used in these phrases are explained as follows: The variable \{{\em item}\} is replaced with a domain-specific term: we use ``product'' for Amazon datasets, ``business'' for Yelp, ``news'' for MIND, and ``recipe'' for Food. The variable \{{\em user\_item\_history}\} refers to the target user's interaction history, which includes the title, category, and description of each item. The variable \{{\em candidate\_items}\} represents a set of 10 candidate items. Unless otherwise stated, the \texttt{system} role is set as \texttt{default\_sys\_inst}. An example of the \textbf{Rephrase} prompt, which is introduced in the next section, is shown in Figure~\ref{list:rephrase}\footnote{For more details on other prompts, please see Section A in the supplementary material.}.

\begin{table}[htbp]
\centering
\caption{Standardized phrasing used in prompts}
\label{table:prompt_components}
\scalebox{0.845}{
\begin{tabular}{p{1.15\linewidth}} 
\toprule
\texttt{default\_sys\_inst}: ``{\em You are an AI assistant that helps people find information}'' \\ 
\midrule

\texttt{user\_info}: ``{\em \# Requirements: you must rank candidate \{item\}s that will be provided below to the target user for recommendation. \# Observation: \{user\_item\_history\}}'' \\ 
\midrule

\texttt{candidates\_info}: ``{\em Based on the above user information, please rank the candidate \{item\}s that align closely with the user's preferences. If item IDs [101, 102, 103] are sorted as first 102, second 103, third 101 in order, present your response in the format below: [3,1,2]. In this example, the number of candidate \{item\}s was 3, but next, 10 \{item\}s will be provided, so you SHOULD sort 10 \{item\}s. \# Candidates \{item\}s: \{candidate\_items\}}'' \\ 
\midrule

\texttt{task\_inst}: ``{\em I here repeat candidate instruction. Please rank 10 candidate \{item\} IDs as follows: If \{item\} IDs [101, 102, 103] are sorted as first 102, second 103, third 101 in order, present your response in the format below: [3,1,2]. \# Additional constraints: - You ONLY rank the given Candidate \{item\} IDs. - Do not explain the reason and include any other words.}'' \\ 
\midrule

\texttt{final\_inst}: ``{\em Thank you! Based on the above conversation, please provide the final answer. \{\texttt{task\_inst}\}}'' \\ 
\midrule

\texttt{preamble}: ``{\em Our final goal is to provide an answer to the following problem. Before tackling this issue,}'' \\
\bottomrule
\end{tabular}
}
\end{table}

\begin{figure}[htbp]
\centering
\begin{lstlisting}[frame=single]
(*@{$t_{\mathrm{sys}}$}@*) = (*@\textcolor{blue}{(default\_sys\_inst)}@*)
------------------------------------------------------------------
(*@{$t_{u1}$@*) = # Requirements: you must rank candidate products that will be provided below to the target user for recommendation. # Observation: {1: {'title': 'The Bourne Legacy Steelbook', 'categories': 'Blu-ray, Movies', 'description': "NOTICE: Polish Release, cover may ...", 'rating': 5.0, 'review': 'awesome movie'}, 2 : (*@\textcolor{blue}{(omitted)}@*) }
(*@\textcolor{blue}{(candidate\_info)}@*)
(*@\textcolor{blue}{(preamble)}@*), rephrase and expand it to help you do better answering. Maintain all information in the original question.
------------------------------------------------------------------
(*@{$t_{a1}$@*) = The task at hand is to evaluate and rank a list of candidate products based on their alignment with the preferences of a target user. The user has previously shown a strong interest in action-packed films, particularly those in the Bourne series, as well as comedies with high ratings. The ranking should reflect the following considerations: 1. The presence of action elements, particularly in the context of the Bourne series. 2. (*@\textcolor{blue}{(omitted)}@*)
Now, let's proceed to rank the candidate products based on these criteria. (*@\textcolor{blue}{(omitted)}@*) 
Based on the analysis of these products, the final ranking will be provided in the specified format. Now, let's rank the products. ### Ranked List:1. **The Bourne Identity** (Action) - Strong alignment with the user's interest in the Bourne series.(*@\textcolor{blue}{(omitted)}@*)10. **Cooking French, Vol. 2** (Special Interests) - Least relevant to the user's interests. ### Final Output:[4, 9, 8, 6, 1, 3, 2, 7, 10, 5]
------------------------------------------------------------------
(*@{$t_{u2}$@*) = (*@\textcolor{blue}{(final\_inst)}@*)
------------------------------------------------------------------
(*@{$t_{a2}$@*) = [4, 9, 8, 6, 1, 3, 2, 7, 10, 5]
\end{lstlisting}
\caption{Multi-turn conversation $[t_{\mathrm{sys}}, t_{u1}, t_{a1}, t_{u2}]$ used in the Rephrase prompt as input to the LLM, with output text $t_{a2}$.}
\label{list:rephrase}
\end{figure}

\subsubsection{Non-Conversational Prompts}
We first introduce prompts that use a single instruction message $t_{u1}$ from the \texttt{user} role without a conversational format, as shown in Table~\ref{table:non_conversational}.

\begin{table}[htbp]
\centering
\caption{Non-conversational prompts.}
\label{table:non_conversational}
\scalebox{0.851}{
\begin{tabular}{p{1.14\linewidth}}
\toprule
\textbf{Baseline}~\cite{DBLP:journals/corr/23_is_chatgpt,DBLP:conf/recsys/DaiSZYSXS0X23_listwise}: We use this minimum component structure as the baseline: $t_{u1} = \texttt{user\_info} + \texttt{candidates\_info} + \texttt{task\_inst}$. \\
\midrule

\textbf{Emotion}~\cite{emotional_li2023largelanguagemodelsunderstand}: Add an emotionally expressive text to encourage more careful processing: $t_{u1} = \textbf{Baseline} +$ {\em ``I want to say this before anything else. These results will have a significant impact on my career. Therefore, please think this seriously.''} \\
\midrule

\textbf{Re-Reading}~\cite{DBLP:conf/emnlp/XuT0XXLLM24_re-reading}: Encourage the LLM to read the prompt twice: $t_{u1} = \textbf{Baseline} +$ {\em ``Read the question again:''} $+ \textbf{Baseline}$.\\
\midrule

\textbf{Both-Inst}~\cite{DBLP:conf/nldb/ClavieCNSB23_bothinst}: Split the instructions: $t_{\text{sys}} = \texttt{default\_sys\_inst} + \texttt{user\_info}$, and $t_{u1} = \texttt{candidates\_info} + \texttt{task\_inst}$. \\
\midrule

\textbf{Recency-Focused}~\cite{DBLP:conf/ecir/HouZLLXMZ24_sort}: Highlight the most recent interaction: $t_{u1} = \texttt{user\_info} +$ {\em ``Note that my most recently \{item\} is \{latest item from user\_item\_history\}.''} $+ \texttt{candidates\_info} + \texttt{task\_inst}$. \\
\midrule

\textbf{RolePlay-User}~\cite{DBLP:conf/naacl/KongZCLQSZWD24_roleplaying,DBLP:conf/acl/WangPQLZWGGN00024_rolellm,DBLP:conf/emnlp/TsengHHCHMC24_twotale_persona,DBLP:journals/corr/persona_survey_tmlr}: Ask the LLM to act as the target user: $t_{\text{sys}} = $ {\em ``You are an AI assistant that pretends to be a person who has interacted with the following items. \# Logs:\{user\_item\_history\}''}, $t_{u1} = \texttt{candidates\_info}' + \texttt{task\_inst}$. In \texttt{candidates\_info}', the first sentence of \texttt{candidates\_info} is replaced with: {\em ``Please rank the candidate \{item\}s that align closely with your preferences.''} \\
\midrule

\textbf{RolePlay-Expert}~\cite{DBLP:conf/naacl/KongZCLQSZWD24_roleplaying,DBLP:conf/acl/WangPQLZWGGN00024_rolellm,DBLP:conf/emnlp/TsengHHCHMC24_twotale_persona,DBLP:journals/corr/persona_survey_tmlr}: Replace the \texttt{system} message with: $t_{\text{sys}} = ${\em ``You are an AI expert in \{dataset\} recommendation.''}; $t_{u1}=$\textbf{Baseline}. In this paper, we set \{{\em dataset}\} to ``business'' for Yelp, ``news'' for MIND, ``recipe'' for Food, and the dataset name itself for others (e.g., the Music category of the Amazon dataset is ``music'').\\
\midrule

\textbf{RolePlay-Frederick}~\cite{DBLP:conf/nldb/ClavieCNSB23_bothinst}: Add a name to the \texttt{system} role: $t_{\text{sys}} = $ {\em ``You are Frederick.''} $+ \texttt{default\_sys\_inst}$; $t_{u1}=$\textbf{Baseline}. \\
\midrule

\textbf{Summarize-Item}~\cite{DBLP:conf/recsys/AcharyaSO23_itemgenerate,DBLP:conf/www/ZhengCQZ024_block_summary}: The format is the same as \textbf{Baseline}, except that item descriptions in \{{\em user\_item\_history}\} and \{{\em candidate\_items}\} are summarized in advance using the following prompt: {\em ``We have the following item details in the \{dataset\} domain: \# Item Information: \{item\_info\} Please summarize this item information, ensuring to include key features and attributes to enhance clarity and understanding within 10--200 words.''} This helps reduce prompt length when item descriptions are long or noisy. \\
\bottomrule
\end{tabular}
}
\end{table}

\subsubsection{Conversational Prompts}
We next introduce multi-turn conversational prompts. In this format, the \texttt{user} role first sends a message $t_{u1}$, and the \texttt{assistant} role replies with $t_{a1}$. Then, a second \texttt{user} role message $t_{u2}$ is used to request the final output.

We begin with prompts where the first \texttt{user} message is set as $t_{u1} = \texttt{user\_info} + \texttt{candidates\_info} + \texttt{middle\_inst}$. Unless otherwise stated, $t_{u2}$ is set to \texttt{final\_inst}. Table~\ref{table:middle_prompts} summarizes the variations used for \texttt{middle\_inst}.

\begin{table}[htbp]
\centering
\caption{\texttt{middle\_inst} used in conversational prompts.}
\label{table:middle_prompts}
\scalebox{0.851}{
\begin{tabular}{p{1.14\linewidth}}
\toprule
\textbf{Explain}~\cite{DBLP:conf/nips/Wei0SBIXCLZ22_CoT,DBLP:conf/dasfaa/LuoCZLC24_exprec}:  
Explain the reason of the item order. $\texttt{middle\_inst} = $ {\em ``Please also include the reason for arranging them in that order.''} \\
\midrule

\textbf{Mock}~\cite{DBLP:conf/nldb/ClavieCNSB23_bothinst}:  
Confirm understanding using $\texttt{middle\_inst} = $ {\em ``Do you understand?''} and set the next reply as $t_{a1} = ${\em ``Yes! I understand. I am ready to answer your question.''} \\
\bottomrule
\end{tabular}
}
\end{table}

Table~\ref{table:preamble_prompts} summarizes prompts that include a \texttt{preamble} in the form $t_{u1} = $ \texttt{user\_info} $+$ \texttt{candidates\_info} $+$ \texttt{preamble} $+$ \texttt{post\_inst}, and presents the variations used for \texttt{post\_inst}.

\begin{table}[htbp]
\centering
\caption{\texttt{post\_inst} used in conversational prompts.}
\label{table:preamble_prompts}
\scalebox{0.851}{
\begin{tabular}{p{1.14\linewidth}}
\toprule
\textbf{Step-Back}~\cite{DBLP:conf/iclr/ZhengMCCCLZ24_stepback}:  
Consider background knowledge. $\texttt{post\_inst} = $ {\em ``please consider the principles and theories behind this question.''} \\
\midrule

\textbf{ReAct}~\cite{DBLP:conf/iclr/YaoZYDSN023_react}:  
Follow explicitly specified steps for reasoning. $\texttt{post\_inst} = $ {\em ``please follow this format to proceed step by step with *Observation*, *Thought*, and *Action*: Observation: Observe the user's history and preferences. Thought: Infer the user's tastes or tendencies from the observation. Action: Choose one candidate item and examine its characteristics. Observation: Observe the characteristics of that item. Thought: Consider whether the item matches the user's preferences. (Repeat for multiple items if necessary) Finally, provide your *Answer*.''} \\
\midrule

\textbf{Rephrase}~\cite{DBLP:journals/corr/abs-2311-04205_rephrase}:  
Rewrite the input for better understanding. $\texttt{post\_inst} = $ {\em ``rephrase and expand it to help you do better answering. Maintain all information in the original question.''} \\
\midrule

\textbf{Echo}~\cite{DBLP:conf/naacl/echo_MekalaRS24}:  
Repeat the input before answering. $\texttt{post\_inst} = $ {\em ``let's repeat the question''}. We then fix $t_{a2}$ as $t_{a1}$. \\
\midrule

\textbf{Summarize-User}~\cite{DBLP:journals/corr/23_summary,DBLP:journals/corr/abs-2402-15623_user_profile,DBLP:conf/www/ZhengCQZ024_block_summary}:  
Describe the user based on their behavior. $\texttt{post\_inst} = $ {\em ``it is important to express who the target user is. Please describe the user in 100--500 words in a way that contributes to the final objective.''} \\
\midrule

\textbf{Generate-Item}~\cite{DBLP:conf/recsys/Kusano24_generate_item,DBLP:journals/corr/abs-2502-02061_reason4rec,DBLP:conf/ecir/JiLXHGTZ24_genrec}:  
Describe items the user might like. $\texttt{post\_inst} = $ {\em ``please describe the \{item\} features that the user might like.''} \\
\midrule

\textbf{Reuse-Item} (Original):  
Use the same instruction as \textbf{Generate-Item}, but remove the most recent item from \{{\em user\_item\_history}\}. We fix $t_{a1}$ as the text of the most recent item. \\
\bottomrule
\end{tabular}
}
\end{table}

Table~\ref{table:initial_response_prompts} presents three prompts that control the initial response of the LLM. Each prompt begins with a \texttt{user} message set as $t_{u1} = \texttt{user\_info} + \texttt{candidates\_info}$, followed by a fixed \texttt{assistant} reply $t_{a1}$. After this, we send $t_{u2} = $ {\em ``OK! Go ahead!''} to prompt the LLM to generate a full response $t_{a2}$. Finally, $t_{u3} = \texttt{final\_inst}$ is used to request the ranking result.

\begin{table}[htbp]
\centering
\caption{Fixed initial responses ($t_{a1}$).}
\label{table:initial_response_prompts}
\scalebox{0.851}{
\begin{tabular}{p{1.14\linewidth}}
\toprule
\textbf{Step-by-Step}~\cite{DBLP:conf/nips/KojimaGRMI22_zscot}:  
Begin with a short instruction for step-by-step thinking. $t_{a1} = $ {\em ``Let's think step by step.''} \\
\midrule

\textbf{Deep-Breath}~\cite{DBLP:conf/iclr/Yang0LLLZC24_breath}:  
Insert a calming phrase to promote careful reasoning. $t_{a1} = $ {\em ``take a deep breath and work on this problem step by step.''} \\
\midrule

\textbf{Plan-Solve}~\cite{DBLP:conf/acl/WangXLHLLL23_plan}:  
Structure the response in three stages: understanding, planning, and solving. $t_{a1} = $ {\em ``Let's first understand the problem and devise a plan to solve the problem. Then, let's carry out the plan and solve the problem step by step.''} \\
\bottomrule
\end{tabular}
}
\end{table}

\subsection{Evaluation Metrics and Output Formatting}
\label{subsubsec:evaluation}

In the numerical experiments, each prompt produced a ranking of 10 candidate items. We evaluated the ranking performance using two standard metrics: Hit@$3$ and nDCG@$3$\footnote{All tasks in this paper fix the same setting to re-rank 10 candidate items, so we use $k=3$ for evaluating nDCG@$k$ and Hit@$k$. Another reason is that some LLMs do not follow the task instruction to re-rank all items and return only a partial ranking. For \texttt{gpt-4o-mini} in the experiments, the probability of this behavior was usually below 2\% when $k=3$, but exceeded 3\% in most cases when $k=5$ (see Section~\ref{subsec:error_analysis} in this paper and Section C in the supplementary material for details).}.

In some cases, LLMs returned only a partial ranking, including just a few top-ranked items. We preserved the order of these items and randomly ranked the rest to complete a full list of 10. If an LLM failed to return any ranking, we retried up to five times. After five failures, we used a fully random ranking.

\section{Performance Evaluation for Prompts}
\label{sec:prompt-level}

This section evaluates the effectiveness of the 21 prompt types introduced in Section~\ref{subsec:prompt} in terms of recommendation accuracy. The goal is to understand which prompts are effective (RQ1), under what conditions (RQ2), and why (RQ3). To make large-scale inference feasible within budget constraints, we conduct experiments using the five low-cost LLMs listed in Table~\ref{table:llms}.

\noindent ~~~~ \textbf{RQ1}: Which prompts significantly outperform the baseline?

\noindent ~~~~ \textbf{RQ2}: Are there any prompts that show stable effectiveness across different LLMs?

\noindent ~~~~ \textbf{RQ3}: Why do some prompts result in lower accuracy?

\subsection{RQ1: Model-Wise Effectiveness}
\label{subsec:wilcoxon}

We investigate whether each prompt leads to higher recommendation accuracy compared to the \textbf{Baseline}, analyzing the results separately for each LLM. This analysis helps identify which prompts are effective for specific LLMs.

\subsubsection{Evaluation Method}

For each LLM, we conducted a hypothesis test to determine whether a prompt performed better than the \textbf{Baseline} using a one-sided Wilcoxon signed-rank test\footnote{The Wilcoxon test is a non-parametric statistical method for comparing paired values. For each user $u$, we calculate the difference $e_u = s^p_u - s^{\textbf{Baseline}}_u$, where $s^p_u$ is the score for prompt $p$. Users with $e_u = 0$ are excluded. We rank the absolute values $|e_u|$ and compute the test statistic from the sum of ranks where $e_u > 0$. The one-sided $p$-value is calculated using a normal approximation.}. We also calculated the relative improvement of each prompt using $r^p = (m^p - m^{\textbf{Baseline}})/m^{\textbf{Baseline}}$, where $m^p$ is the average score for prompt $p$. These evaluations used all 1,600 users, as well as specific subsets, such as individual datasets (e.g., 200 users from the Movies dataset) or specific user types (e.g., heavy users, $n = 800$).

Table~\ref{table:41_ndcg_scores} shows the average nDCG@$3$ scores for each prompt and dataset using \texttt{gpt-4.1-mini}. The last five columns report the overall scores for all users evaluated with the other low-cost LLMs. Relative improvements are summarized in Table~\ref{table:all_metric_improvement}\footnote{For results with other LLMs, datasets, and Hit@$3$, please refer to Section B in the supplementary material.}.

\begin{table*}[htbp]
\caption{nDCG@$3$ for \texttt{gpt-4.1-mini} and other LLMs. The best prompt in each column is underlined. Prompts significantly better or worse than \textbf{Baseline} ($p < 0.05$) are marked with $*$ or $\bigtriangledown$, respectively.}
\label{table:41_ndcg_scores}
\centering
\scalebox{0.78}{
\begin{tabular}{c|cccccccc|cc||ccccc}
\toprule
& Music & Movie & Groceries & Clothes & Book & Yelp & News & Food & Light & Heavy & 4.1-mini & llama3.3 & 4o-mini & phi-4 & nova-lite  \\
\midrule
Baseline & $0.610$ & $0.516$ & $0.443$ & $0.408$ & $0.585$ & $0.512$ & $0.262$ & $0.328$ & $0.460$ & $0.456$ & $0.458$ & $0.439$ & $0.432$ & $0.417$ & $0.389$ \\
Emotion & $0.608$ & \cellcolor{nlow}$0.491^{\bigtriangledown}$ & \cellcolor{nslow}$0.438$ & $0.411$ & $0.582$ & \cellcolor{nslow}$0.503$ & \cellcolor{plow}$0.271$ & $0.327$ & $0.461$ & \cellcolor{nslow}$0.447$ & $0.454$ & $0.442$ & \cellcolor{pslow}$0.440$ & $0.417$ & \cellcolor{nlow}$0.376^{\bigtriangledown}$ \\
Re-Reading & \cellcolor{nhigh}$0.567^{\bigtriangledown}$ & \cellcolor{nmiddle}$0.487$ & \cellcolor{nmiddle}$0.420$ & \cellcolor{nslow}$0.400$ & \cellcolor{nmiddle}$0.553^{\bigtriangledown}$ & \cellcolor{nshigh}$0.421^{\bigtriangledown}$ & \cellcolor{plow}$0.275$ & \cellcolor{nhigh}$0.304$ & \cellcolor{nlow}$0.444$ & \cellcolor{nhigh}$0.412^{\bigtriangledown}$ & \cellcolor{nmiddle}$0.428^{\bigtriangledown}$ & \cellcolor{phigh}$0.472^{*}$ & \cellcolor{nmiddle}$0.402^{\bigtriangledown}$ & \cellcolor{nslow}$0.409$ & \cellcolor{pslow}$0.393$ \\
Both-Inst & $0.609$ & $0.513$ & $0.443$ & \cellcolor{nslow}$0.398$ & \cellcolor{nlow}$0.566$ & \cellcolor{pmiddle}$0.544^{*}$ & \cellcolor{nshigh}$0.226^{\bigtriangledown}$ & \cellcolor{nlow}$0.316$ & $0.459$ & \cellcolor{nslow}$0.444$ & \cellcolor{nslow}$0.452$ & $0.437$ & $0.430$ & \cellcolor{nlow}$0.403$ & \cellcolor{pslow}$0.396$ \\
Recency-Focused & \cellcolor{nslow}$0.601$ & $0.519$ & \cellcolor{pmiddle}$0.466$ & \cellcolor{pslow}$0.420$ & \cellcolor{pslow}$0.591$ & \cellcolor{pslow}$0.526$ & \cellcolor{nmiddle}$0.245$ & \cellcolor{pshigh}$\underline{0.361}$ & \cellcolor{plow}$0.478$ & $0.455$ & \cellcolor{pslow}$0.466$ & \cellcolor{pmiddle}$0.470^{*}$ & \cellcolor{pslow}$0.443$ & \cellcolor{nslow}$0.409$ & \cellcolor{pslow}$0.394$ \\
RolePlay-User & \cellcolor{nshigh}$0.540^{\bigtriangledown}$ & \cellcolor{nmiddle}$0.484$ & \cellcolor{nshigh}$0.372^{\bigtriangledown}$ & \cellcolor{nshigh}$0.325^{\bigtriangledown}$ & \cellcolor{nhigh}$0.533^{\bigtriangledown}$ & \cellcolor{nslow}$0.500$ & \cellcolor{nshigh}$0.195^{\bigtriangledown}$ & \cellcolor{nlow}$0.317$ & \cellcolor{nhigh}$0.420^{\bigtriangledown}$ & \cellcolor{nshigh}$0.397^{\bigtriangledown}$ & \cellcolor{nshigh}$0.408^{\bigtriangledown}$ & \cellcolor{pslow}$0.447$ & \cellcolor{nlow}$0.411^{\bigtriangledown}$ & \cellcolor{nshigh}$0.375^{\bigtriangledown}$ & $0.391$ \\
RolePlay-Expert & \cellcolor{nslow}$0.592$ & \cellcolor{nslow}$0.507$ & $0.441$ & \cellcolor{nmiddle}$0.380^{\bigtriangledown}$ & $0.587$ & $0.515$ & $0.260$ & \cellcolor{nhigh}$0.305$ & \cellcolor{nslow}$0.450$ & \cellcolor{nslow}$0.446$ & \cellcolor{nslow}$0.448^{\bigtriangledown}$ & $0.440$ & $0.431$ & $0.415$ & \cellcolor{nslow}$0.383$ \\
RolePlay-Frederick & \cellcolor{nlow}$0.588^{\bigtriangledown}$ & \cellcolor{nslow}$0.505$ & $0.441$ & \cellcolor{nslow}$0.403$ & \cellcolor{nslow}$0.569$ & $0.511$ & $0.261$ & \cellcolor{nhigh}$0.303$ & \cellcolor{nslow}$0.451$ & \cellcolor{nslow}$0.445^{\bigtriangledown}$ & \cellcolor{nslow}$0.448^{\bigtriangledown}$ & $0.437$ & \cellcolor{nslow}$0.426$ & $0.418$ & \cellcolor{nslow}$0.379$ \\
Summarize-Item & \cellcolor{nshigh}$0.536^{\bigtriangledown}$ & \cellcolor{nlow}$0.495$ & \cellcolor{nhigh}$0.401$ & \cellcolor{plow}$0.421$ & \cellcolor{nhigh}$0.536^{\bigtriangledown}$ & \cellcolor{nmiddle}$0.481$ & \cellcolor{nshigh}$0.234$ & \cellcolor{nslow}$0.318$ & \cellcolor{nmiddle}$0.431^{\bigtriangledown}$ & \cellcolor{nmiddle}$0.424^{\bigtriangledown}$ & \cellcolor{nmiddle}$0.428^{\bigtriangledown}$ & \cellcolor{pslow}$0.446$ & \cellcolor{nslow}$0.427$ & \cellcolor{nshigh}$0.371^{\bigtriangledown}$ & \cellcolor{nslow}$0.378$ \\
Step-Back & \cellcolor{pmiddle}$0.642$ & \cellcolor{phigh}$0.561^{*}$ & \cellcolor{pmiddle}$0.469$ & \cellcolor{phigh}$0.437$ & \cellcolor{pslow}$0.594$ & \cellcolor{pslow}$0.521$ & \cellcolor{nmiddle}$0.246$ & \cellcolor{plow}$0.340$ & \cellcolor{plow}$0.480^{*}$ & \cellcolor{plow}$0.473$ & \cellcolor{plow}$0.476^{*}$ & \cellcolor{phigh}$0.481^{*}$ & \cellcolor{pmiddle}$0.460^{*}$ & \cellcolor{plow}$\underline{0.436}$ & \cellcolor{nshigh}$0.348^{\bigtriangledown}$ \\
ReAct & \cellcolor{plow}$0.632$ & \cellcolor{phigh}$\underline{0.566^{*}}$ & \cellcolor{pshigh}$\underline{0.511^{*}}$ & \cellcolor{phigh}$\underline{0.448}$ & \cellcolor{plow}$0.608$ & \cellcolor{phigh}$\underline{0.548^{*}}$ & \cellcolor{nmiddle}$0.248$ & \cellcolor{pslow}$0.334$ & \cellcolor{plow}$0.479^{*}$ & \cellcolor{phigh}$\underline{0.494^{*}}$ & \cellcolor{pmiddle}$\underline{0.487^{*}}$ & \cellcolor{pshigh}$\underline{0.495^{*}}$ & \cellcolor{nshigh}$0.369^{\bigtriangledown}$ & $0.413$ & \cellcolor{nmiddle}$0.369$ \\
Rephrase & \cellcolor{pslow}$0.626$ & \cellcolor{phigh}$0.564^{*}$ & \cellcolor{plow}$0.461$ & \cellcolor{phigh}$0.437$ & \cellcolor{phigh}$\underline{0.627^{*}}$ & \cellcolor{pslow}$0.517$ & \cellcolor{nlow}$0.253$ & \cellcolor{phigh}$0.359$ & \cellcolor{pmiddle}$\underline{0.485^{*}}$ & \cellcolor{plow}$0.476^{*}$ & \cellcolor{plow}$0.481^{*}$ & \cellcolor{pmiddle}$0.467^{*}$ & \cellcolor{plow}$0.445^{*}$ & \cellcolor{pslow}$0.429$ & \cellcolor{pslow}$\underline{0.398}$ \\
Echo & \cellcolor{nslow}$0.594$ & \cellcolor{pslow}$0.525$ & \cellcolor{nlow}$0.426$ & \cellcolor{nmiddle}$0.383$ & \cellcolor{nmiddle}$0.546^{\bigtriangledown}$ & \cellcolor{nshigh}$0.449^{\bigtriangledown}$ & \cellcolor{plow}$0.270$ & \cellcolor{nshigh}$0.282$ & \cellcolor{nmiddle}$0.433^{\bigtriangledown}$ & \cellcolor{nlow}$0.436^{\bigtriangledown}$ & \cellcolor{nmiddle}$0.434^{\bigtriangledown}$ & \cellcolor{phigh}$0.477^{*}$ & \cellcolor{nhigh}$0.397^{\bigtriangledown}$ & $0.420$ & \cellcolor{nshigh}$0.349^{\bigtriangledown}$ \\
Summarize-User & \cellcolor{pmiddle}$\underline{0.646^{*}}$ & \cellcolor{pslow}$0.523$ & \cellcolor{pmiddle}$0.472$ & $0.406$ & $0.580$ & \cellcolor{nslow}$0.506$ & \cellcolor{pslow}$0.266$ & $0.324$ & \cellcolor{plow}$0.480^{*}$ & \cellcolor{nslow}$0.452$ & \cellcolor{pslow}$0.466$ & \cellcolor{pmiddle}$0.467^{*}$ & \cellcolor{plow}$0.453^{*}$ & $0.417$ & \cellcolor{nlow}$0.372^{\bigtriangledown}$ \\
Generate-Item & $0.612$ & $0.515$ & \cellcolor{plow}$0.457$ & \cellcolor{pslow}$0.419$ & $0.581$ & \cellcolor{plow}$0.532$ & \cellcolor{pslow}$0.268$ & \cellcolor{nmiddle}$0.306^{\bigtriangledown}$ & \cellcolor{pslow}$0.467$ & $0.455$ & $0.461$ & $0.438$ & \cellcolor{pslow}$0.445^{*}$ & \cellcolor{nslow}$0.406$ & \cellcolor{nslow}$0.383$ \\
Reuse-Item & $0.606$ & \cellcolor{pmiddle}$0.551^{*}$ & \cellcolor{nmiddle}$0.415$ & \cellcolor{nlow}$0.391$ & \cellcolor{nslow}$0.578$ & \cellcolor{plow}$0.527$ & \cellcolor{nshigh}$0.229^{\bigtriangledown}$ & $0.329$ & $0.456$ & \cellcolor{nslow}$0.451$ & \cellcolor{nslow}$0.453$ & \cellcolor{pmiddle}$0.467^{*}$ & \cellcolor{nslow}$0.427$ & \cellcolor{nhigh}$0.382^{\bigtriangledown}$ & \cellcolor{nlow}$0.370^{\bigtriangledown}$ \\
Explain & \cellcolor{pslow}$0.616$ & $0.514$ & \cellcolor{phigh}$0.476^{*}$ & $0.412$ & \cellcolor{nslow}$0.571$ & \cellcolor{plow}$0.530$ & $0.261$ & $0.326$ & \cellcolor{pslow}$0.466$ & \cellcolor{pslow}$0.461$ & \cellcolor{pslow}$0.463$ & \cellcolor{pmiddle}$0.465^{*}$ & \cellcolor{pmiddle}$\underline{0.460^{*}}$ & $0.415$ & \cellcolor{nslow}$0.379$ \\
Mock & $0.614$ & \cellcolor{nlow}$0.495$ & $0.439$ & \cellcolor{pslow}$0.415$ & $0.580$ & \cellcolor{pslow}$0.521$ & \cellcolor{nshigh}$0.235^{\bigtriangledown}$ & \cellcolor{nslow}$0.320$ & \cellcolor{nslow}$0.455$ & \cellcolor{nslow}$0.450$ & \cellcolor{nslow}$0.452$ & \cellcolor{pslow}$0.451^{*}$ & \cellcolor{nmiddle}$0.408^{\bigtriangledown}$ & \cellcolor{nmiddle}$0.395^{\bigtriangledown}$ & \cellcolor{nlow}$0.373^{\bigtriangledown}$ \\
Step-by-Step & $0.608$ & $0.511$ & \cellcolor{nlow}$0.424$ & \cellcolor{plow}$0.428$ & $0.580$ & $0.509$ & \cellcolor{pmiddle}$0.277$ & $0.326$ & $0.464$ & $0.452$ & $0.458$ & $0.439$ & \cellcolor{nslow}$0.428$ & \cellcolor{nslow}$0.408$ & \cellcolor{nslow}$0.382$ \\
Deep-Breath & $0.615$ & $0.517$ & \cellcolor{nslow}$0.436$ & \cellcolor{plow}$0.428^{*}$ & $0.591$ & \cellcolor{pslow}$0.525$ & \cellcolor{phigh}$\underline{0.286^{*}}$ & \cellcolor{nlow}$0.313$ & $0.463$ & \cellcolor{pslow}$0.465$ & \cellcolor{pslow}$0.464$ & $0.440$ & \cellcolor{nslow}$0.426$ & \cellcolor{nslow}$0.407$ & \cellcolor{nlow}$0.373^{\bigtriangledown}$ \\
Plan-Solve & \cellcolor{nslow}$0.602$ & \cellcolor{nslow}$0.501$ & $0.443$ & \cellcolor{plow}$0.427$ & \cellcolor{nslow}$0.573$ & \cellcolor{pslow}$0.526$ & \cellcolor{nslow}$0.259$ & \cellcolor{nslow}$0.318$ & $0.462$ & \cellcolor{nslow}$0.450$ & $0.456$ & $0.442$ & \cellcolor{nslow}$0.422^{\bigtriangledown}$ & \cellcolor{nlow}$0.402$ & \cellcolor{nhigh}$0.358^{\bigtriangledown}$ \\
\bottomrule
\end{tabular}
}
\end{table*}

\begin{table}[htbp]
\caption{Average relative improvements in nDCG@$3$ and Hit@$3$. Prompts with significant changes are marked with $*$ (better) or $\bigtriangledown$ (worse), as in Table~\ref{table:41_ndcg_scores}. ``Avg'' shows the mean across all LLMs, with parentheses showing the total count of significant differences.}
\label{table:all_metric_improvement}
\centering
\scalebox{0.745}{
\begin{tabular}{c|ccccc|c}
\toprule
 & 4.1-mini & llama3.3 & 4o-mini & phi-4 & nova-lite & Avg \\
\midrule
Rephrase & \cellcolor{plow}$4.5^{*}$ & \cellcolor{pmiddle}$5.9^{*}$ & \cellcolor{pslow}$2.4$ & \cellcolor{pslow}$3.0$ & \cellcolor{pslow}$1.7$ & \cellcolor{plow}$3.5 (5)$ \\
Step-Back & \cellcolor{plow}$4.4^{*}$ & \cellcolor{phigh}$8.9^{*}$ & \cellcolor{pmiddle}$5.6^{*}$ & \cellcolor{plow}$3.6$ & \cellcolor{nshigh}$-11.1^{\bigtriangledown}$ & \cellcolor{pslow}$2.3 (4)$ \\
Explain & \cellcolor{pslow}$1.0$ & \cellcolor{pmiddle}$6.1^{*}$ & \cellcolor{pmiddle}$5.3^{*}$ & $-0.2$ & \cellcolor{nslow}$-2.3$ & \cellcolor{pslow}$2.0 (4)$ \\
Summarize-User & \cellcolor{pslow}$1.2$ & \cellcolor{pmiddle}$5.9^{*}$ & \cellcolor{plow}$4.3^{*}$ & $0.5$ & \cellcolor{nmiddle}$-5.2^{\bigtriangledown}$ & \cellcolor{pslow}$1.3 (2)$ \\
Recency-Focused & $0.9$ & \cellcolor{pmiddle}$6.0^{*}$ & \cellcolor{pslow}$1.4$ & \cellcolor{nslow}$-1.2$ & \cellcolor{pslow}$1.4$ & \cellcolor{pslow}$1.7 (2)$ \\
ReAct & \cellcolor{pmiddle}$5.8^{*}$ & \cellcolor{pshigh}$12.0^{*}$ & \cellcolor{nshigh}$-13.8^{\bigtriangledown}$ & $-0.2$ & \cellcolor{nmiddle}$-5.3$ & $-0.3 (1)$ \\
Generate-Item & $0.4$ & $-0.2$ & \cellcolor{pslow}$2.3$ & \cellcolor{nslow}$-2.3$ & $-0.9$ & $-0.2 (1)$ \\
Both-Inst & \cellcolor{nslow}$-1.3$ & $-0.6$ & \cellcolor{nslow}$-1.4$ & \cellcolor{nlow}$-3.3$ & \cellcolor{pslow}$1.7$ & $-1.0$ \\
Step-by-Step & $-0.3$ & $-0.1$ & \cellcolor{nslow}$-1.7$ & \cellcolor{nslow}$-1.6$ & \cellcolor{nslow}$-1.9$ & \cellcolor{nslow}$-1.1$ \\
RolePlay-Expert & \cellcolor{nslow}$-1.6$ & $0.4$ & $-0.9$ & $-0.6$ & $-1.0$ & $-0.8 (-1)$ \\
Emotion & $-1.0$ & \cellcolor{pslow}$1.1$ & \cellcolor{pslow}$1.5$ & $0.5$ & \cellcolor{nlow}$-3.5^{\bigtriangledown}$ & $-0.3 (-2)$ \\
Re-Reading & \cellcolor{nmiddle}$-6.1^{\bigtriangledown}$ & \cellcolor{phigh}$7.6^{*}$ & \cellcolor{nhigh}$-7.6^{\bigtriangledown}$ & \cellcolor{nslow}$-1.4$ & \cellcolor{pslow}$1.9$ & \cellcolor{nslow}$-1.1 (-2)$ \\
Reuse-Item & \cellcolor{nslow}$-1.3$ & \cellcolor{pmiddle}$5.6^{*}$ & \cellcolor{nslow}$-1.7$ & \cellcolor{nhigh}$-7.7^{\bigtriangledown}$ & \cellcolor{nmiddle}$-5.0^{\bigtriangledown}$ & \cellcolor{nslow}$-2.0 (-2)$ \\
Deep-Breath & \cellcolor{pslow}$1.3$ & $0.1$ & \cellcolor{nslow}$-1.8$ & \cellcolor{nslow}$-2.2$ & \cellcolor{nlow}$-4.0^{\bigtriangledown}$ & \cellcolor{nslow}$-1.3 (-2)$ \\
RolePlay-Frederick & \cellcolor{nslow}$-2.0$ & $-0.4$ & \cellcolor{nslow}$-1.9$ & $0.5$ & \cellcolor{nslow}$-2.2$ & \cellcolor{nslow}$-1.2 (-2)$ \\
Plan-Solve & $-0.6$ & $0.9$ & \cellcolor{nslow}$-2.3$ & \cellcolor{nslow}$-3.0$ & \cellcolor{nhigh}$-8.3^{\bigtriangledown}$ & \cellcolor{nslow}$-2.7 (-3)$ \\
Summarize-Item & \cellcolor{nmiddle}$-5.8^{\bigtriangledown}$ & \cellcolor{pslow}$1.1$ & \cellcolor{nslow}$-1.8$ & \cellcolor{nhigh}$-9.6^{\bigtriangledown}$ & \cellcolor{nslow}$-2.7$ & \cellcolor{nlow}$-3.8 (-4)$ \\
Echo & \cellcolor{nlow}$-4.9^{\bigtriangledown}$ & \cellcolor{phigh}$8.5^{*}$ & \cellcolor{nhigh}$-8.1^{\bigtriangledown}$ & \cellcolor{pslow}$1.6$ & \cellcolor{nhigh}$-9.0^{\bigtriangledown}$ & \cellcolor{nslow}$-2.4 (-4)$ \\
Mock & \cellcolor{nslow}$-1.1$ & \cellcolor{pslow}$2.2$ & \cellcolor{nmiddle}$-6.5^{\bigtriangledown}$ & \cellcolor{nlow}$-4.6^{\bigtriangledown}$ & \cellcolor{nlow}$-4.1^{\bigtriangledown}$ & \cellcolor{nslow}$-2.8 (-5)$ \\
RolePlay-User & \cellcolor{nshigh}$-10.9^{\bigtriangledown}$ & \cellcolor{pslow}$1.5$ & \cellcolor{nmiddle}$-5.9^{\bigtriangledown}$ & \cellcolor{nhigh}$-9.7^{\bigtriangledown}$ & \cellcolor{pslow}$1.1$ & \cellcolor{nlow}$-4.8 (-6)$ \\
\bottomrule
\end{tabular}
}
\end{table}

\subsubsection{Discussion}
\label{subsubsec:wilcoxon_discussion}

We summarize key findings from the statistical analysis across five cost-efficient LLMs.

\underline{Stable prompts across datasets and LLMs.}
Some prompts that significantly improved performance over the \textbf{Baseline} in \texttt{gpt-4.1-mini} (marked with $*$ in Table~\ref{table:41_ndcg_scores}) also maintained their effectiveness across different datasets and user types. Importantly, none of them showed a clear decrease in performance. Similar trends were observed with other LLMs such as \texttt{llama3.3-70b} and \texttt{gpt-4o-mini}. In particular, the prompts \textbf{Rephrase}, \textbf{Step-Back}, \textbf{Explain}, \textbf{Recency-Focused}, and \textbf{Summarize-User} showed higher accuracy from Table~\ref{table:all_metric_improvement}.

\underline{Smaller effects in low-performing LLMs.}
In results from \texttt{phi-4} and \texttt{amazon-nova-lite}, no prompt showed significant improvement over the \textbf{Baseline}. These models also showed lower accuracy compared to other LLMs. One possible reason is that they may have limited ability to understand detailed prompt instructions. As a result, changes in prompt design had only a small effect.

\underline{Performance variation depending on LLMs.}
Some prompts showed different results depending on the LLM. For example, \textbf{Re-Reading} and \textbf{Echo} improved accuracy in \texttt{llama3.3-70b}, but reduced it in \texttt{gpt-4.1-mini} and \texttt{gpt-4o-mini}. These prompts involve repeated content, which may confuse some models or make it harder to focus on relevant information. Interestingly, \textbf{ReAct} showed the highest accuracy in both \texttt{gpt-4.1-mini} and \texttt{llama3.3-70b}, but performed the worst in \texttt{gpt-4o-mini}.

\underline{Limited effects of NLP-style prompts.}
Prompts commonly used in NLP, such as \textbf{RolePlay}, \textbf{Mock}, \textbf{Plan-Solve}, \textbf{Deep-Breath}, \textbf{Emotion}, and \textbf{Step-by-Step}, did not improve recommendation accuracy. This suggests that such techniques may be less suitable for ranking tasks, where capturing user preferences is essential.

\subsection{RQ2: Model-Agnostic Effectiveness}
\label{subsec:lmem}

We use a linear mixed-effects model (LMEM)\footnote{\url{https://www.statsmodels.org/stable/mixed_linear.html}} to examine how each prompt is effective while controlling for factors such as the LLM, user, and evaluation metric. This method is similar to linear regression but also captures variation across users and LLMs. The model helps identify prompts that perform well under various conditions.

While the previous analysis in Section~\ref{subsec:wilcoxon} showed that performance varied depending on the LLM, this analysis aims to find prompts that are effective regardless of the LLM or dataset. LMEM is suitable for this purpose because it adjusts for differences across users, LLMs, and metrics when evaluating each prompt.

\subsubsection{Evaluation Method}

To evaluate prompt performance across different LLMs and users, we used the following LMEM:
\[
\texttt{acc}_{i} = \beta_0 + \sum_{j=1}^{P} \beta_j \cdot I(\texttt{prompt}_i = j) + \gamma_{\texttt{user}(i)} + \gamma_{\texttt{llm}(i)} + \gamma_{\texttt{metric}(i)} + \varepsilon_i,
\]
where $\texttt{acc}_{i}$ is the accuracy for observation $i$, $I(\cdot)$ is an indicator function, $\beta_j$ is the fixed effect of prompt $j$, and $\gamma_{\cdot}$ are random effects. We tested each $\beta_j$ using a Wald test. If $\beta_j > 0$ and $p < 0.05$, we regard the prompt as significantly better than the \textbf{Baseline}, which serves as the reference.

\begin{table}[htbp]
\caption{Prompts that showed statistically significant positive or negative effects by LMEM compared to the \textbf{Baseline}.}
\label{table:lmem}
\centering
\scalebox{0.82}{
\begin{tabular}{p{0.07\textwidth}|p{0.12\textwidth}|p{0.32\textwidth}}
\toprule
Dataset & Positive & Negative \\
\midrule
Music &  & Summarize-Item, RolePlay-User, Re-Reading \\
Movie & ReAct, Rephrase &  \\
Groceries & ReAct & RolePlay-User \\
Clothes &  & RolePlay-User \\
Book & Rephrase & RolePlay-User, Summarize-Item \\
Yelp &  & Re-Reading, Echo \\
News &  & RolePlay-User \\
Food &  & Echo \\
\midrule
Light & Rephrase & RolePlay-User, Summarize-Item, Echo \\
Heavy & ReAct & Re-Reading, RolePlay-User, Summarize-Item \\
\midrule
ALL & ReAct, Rephrase, Step-Back & Re-Reading, RolePlay-User, Summarize-Item, Echo \\
\bottomrule
\end{tabular}
}
\end{table}
\subsubsection{Discussion}
\label{subsubsec:lmem_discussion}

According to Table~\ref{table:lmem}, \textbf{ReAct}, \textbf{Rephrase}, and \textbf{Step-Back} performed significantly better across all datasets and LLMs. These prompts also achieved higher accuracy in \texttt{gpt-4.1-mini} and \texttt{llama3.3-70b} (see Table~\ref{table:41_ndcg_scores}). Since these two LLMs showed relatively strong performance among the low-cost models, their results may have exerted a stronger influence on the LMEM estimates.

Prompts such as \textbf{Re-Reading}, \textbf{RolePlay-User}, \textbf{Summarize-Item}, and \textbf{Echo} showed lower accuracy in the LMEM analysis. As shown in Table~\ref{table:41_ndcg_scores}, \textbf{RolePlay-User} and \textbf{Summarize-Item} did not improve performance in any LLM, while \textbf{Re-Reading} and \textbf{Echo} were effective only in \texttt{llama3.3-70b} and showed negative effects in others.

\subsection{RQ3: Error Analysis}
\label{subsec:error_analysis}
This section examines why some prompts led to lower accuracy. As described in Section~\ref{subsubsec:evaluation}, if an LLM failed to return any ranking and all five retries also failed, we completed it by random ranking. The left values in Table~\ref{table:partial_ranking_3_0} show these failure rates.

\begin{table}[htbp]
\caption{Percentages of outputs with failures (left) and those containing three or fewer ranked items (right).}
\label{table:partial_ranking_3_0}
\centering
\scalebox{0.8}{
\begin{tabular}{c|ccccc}
\toprule
 & 4.1-mini & llama3.3 & 4o-mini & phi-4 & nova-lite \\
\midrule
Baseline & $0 / 0$ & $0 / 0$ & $0.4 / 1.7$ & $\textbf{16.6} / \textbf{12.5}$ & $0.1 / 0$ \\
Emotion & $0 / 0$ & $0 / 0$ & $0 / 1.0$ & $\textbf{17.0} / \textbf{12.6}$ & $0 / 0$ \\
Re-Reading & $0 / 0$ & $0 / 0$ & $0.4 / 2.1$ & $\textbf{46.1} / \textbf{26.1}$ & $0 / 0$ \\
Both-Inst & $0 / 0$ & $0 / 0$ & $0.4 / 1.0$ & $\textbf{16.4} / \textbf{12.5}$ & $0.1 / 0$ \\
Recency-Focused & $0 / 0$ & $0 / 0$ & $0.8 / 2.1$ & $\textbf{18.0} / \textbf{12.6}$ & $0.1 / 0$ \\
RolePlay-User & $0 / 0$ & $0 / 0$ & $0 / 0.1$ & $\textbf{16.2} / \textbf{12.5}$ & $0 / 0$ \\
RolePlay-Expert & $0 / 0$ & $0 / 0$ & $0.4 / 1.3$ & $\textbf{16.6} / \textbf{12.5}$ & $0 / 0$ \\
RolePlay-Frederick & $0 / 0$ & $0 / 0$ & $0.5 / 1.6$ & $\textbf{16.2} / \textbf{12.8}$ & $0.1 / 0$ \\
Summarize-Item & $0 / 0$ & $0 / 0$ & $0 / 0.3$ & $7.9 / 0$ & $0 / 0$ \\
Step-Back & $0 / 0$ & $0 / 0$ & $0 / 1.7$ & $\textbf{35.5} / \textbf{13.6}$ & $\textbf{39.6} / 0.5$ \\
ReAct & $0.1 / 0$ & $0 / 0$ & $1.4 / \textbf{21.5}$ & $\textbf{26.4} / \textbf{15.1}$ & $\textbf{19.1} / 2.9$ \\
Rephrase & $0 / 0$ & $0 / 0$ & $0.1 / 1.8$ & $\textbf{25.0} / \textbf{14.1}$ & $5.1 / 0$ \\
Echo & $0 / 0$ & $0 / 0$ & $0.5 / 2.2$ & $\textbf{44.9} / \textbf{25.0}$ & $0 / 0$ \\
Summarize-User & $0 / 0$ & $0 / 0$ & $0.1 / 0.8$ & $\textbf{18.8} / \textbf{13.0}$ & $1.7 / 0$ \\
Generate-Item & $0 / 0$ & $0 / 0$ & $0.2 / 1.0$ & $\textbf{24.3} / \textbf{12.5}$ & $0.2 / 0$ \\
Reuse-Item & $0 / 0$ & $0 / 0$ & $0 / 2.4$ & $\textbf{19.5} / \textbf{12.6}$ & $0 / 0$ \\
Explain & $0 / 0$ & $0 / 0$ & $0 / 0.2$ & $\textbf{23.1} / \textbf{12.6}$ & $0.1 / 0$ \\
Mock & $0 / 0$ & $0 / 0$ & $0.1 / 1.5$ & $\textbf{16.9} / \textbf{12.5}$ & $0 / 0$ \\
Step-by-Step & $0 / 0$ & $0 / 0$ & $0 / 1.7$ & $\textbf{36.0} / \textbf{13.5}$ & $0.6 / 0$ \\
Deep-Breath & $0 / 0$ & $0 / 0$ & $0 / 1.7$ & $\textbf{39.1} / \textbf{13.3}$ & $0.4 / 0$ \\
Plan-Solve & $0 / 0$ & $0 / 0$ & $0 / 2.4$ & $\textbf{50.9} / \textbf{14.1}$ & $8.4 / 1.9$ \\
\bottomrule
\end{tabular}
}
\end{table}

For \texttt{gpt-4.1-mini} and \texttt{llama3.3-70b}, almost all outputs followed the expected format. For \texttt{gpt-4o-mini}, \textbf{ReAct} showed a failure rate of 1.4\%. In \texttt{phi-4}, all prompts except \textbf{Summarize-Item} had failure rates above 10\%, mostly due to token length limits\footnote{Detailed generated outputs, token lengths per dataset, and rates for outputs with only one or five or fewer items are provided in Section C of the supplementary material.}. In \texttt{amazon-nova-lite}, both \textbf{ReAct} and \textbf{Step-Back} exceeded 10\%. For LLMs other than \texttt{phi-4} and \texttt{amazon-nova-lite}, many failures included responses such as {\em ``the user is interested in the candidate items XX, XX, XX.''}, where the LLM ignored the instruction in \texttt{task\_inst} to return a full ranking\footnotemark[\value{footnote}]. In contrast, failures in \texttt{phi-4} and \texttt{amazon-nova-lite} were often caused by prompts that were too long for the LLMs to process\footnotemark[\value{footnote}]. Since \textbf{Summarize-Item} compresses item descriptions and reduces token length, it resulted in the lowest failure rate in \texttt{phi-4}.

We also examined cases where the LLM returned only a few items, although a full ranking was still constructed by randomly appending the unmentioned ones. Specifically, we calculated the rate at which the output contained three or fewer items. As shown in the right values of Table~\ref{table:partial_ranking_3_0}\footnotemark[\value{footnote}], \textbf{ReAct} in \texttt{gpt-4o-mini} showed a partial output rate of over 20\%. This was the only prompt-model combination in \texttt{gpt-4o-mini} with such a high rate, and this behavior contributed to the lower accuracy observed in Table~\ref{table:41_ndcg_scores}. These results suggest that failing to follow instructions and returning incomplete rankings is a key factor in reduced accuracy.

\subsection{Key Findings of This Section}
\label{subsec:summary}

We summarize the findings from this section. In Section~\ref{subsec:wilcoxon}, we compared recommendation accuracy across different prompts for each LLM and dataset. Prompts such as \textbf{Rephrase}, \textbf{Step-Back}, \textbf{Explain}, \textbf{Recency-Focused}, and \textbf{Summarize-User} improved accuracy under many conditions. Section~\ref{subsec:lmem} extended this analysis using a linear mixed-effects model, which adjusted for differences across users, evaluation metrics, and LLMs. The results confirmed that \textbf{ReAct}, \textbf{Step-Back}, and \textbf{Rephrase} performed well across diverse settings, which are similar to the effective prompts found in Section~\ref{subsec:wilcoxon}. In contrast, as shown in Section~\ref{subsec:error_analysis}, prompts that did not improve accuracy often failed to follow the output format or return complete rankings.

Based on these findings, we conclude that in cost-efficient LLMs such as \texttt{gpt-4.1-mini} and \texttt{llama3.3-70b}, prompts that clarify the task context and goal, such as \textbf{ReAct}, \textbf{Step-Back}, and \textbf{Rephrase}, are especially effective for recommendation.

\section{Toward Improving Accuracy}
\label{sec:enhance}

In Section~\ref{sec:prompt-level}, we identified several prompts that improved recommendation accuracy. In this section, we explore whether additional improvement is possible by combining prompts, applying meta-level techniques, or using more advanced LLMs.

\subsection{Prompt Combinations}

We focus on \textbf{ReAct}, \textbf{Step-Back}, and \textbf{Rephrase}, which were found to be effective in Section~\ref{sec:prompt-level}, and create combinations by using them in sequence. \textbf{ReAct} and \textbf{Step-Back} support structured reasoning, while \textbf{Rephrase} makes the instruction text easier to understand. Based on this, we design two combinations: \textbf{Rephrase $\rightarrow$ ReAct} and \textbf{Rephrase $\rightarrow$ Step-Back}\footnote{Full prompt wordings are provided in Section A of the supplementary material.}.

We also apply meta-prompts that refine outputs based on the LLM's initial reasoning (Table~\ref{table:meta_prompts}). Let $t_{u1}$ be the base prompt (e.g., \textbf{ReAct}), $t_{a1}$ the initial output, and $t_{u3} = \texttt{final\_inst}$. These methods use intermediate reasoning explicitly to improve the final response. We apply \textbf{Self-Refine} and \textbf{Self-Consistency} to \textbf{Rephrase}, \textbf{ReAct}, and \textbf{Step-Back}, resulting in six meta-prompt variants.

\begin{table}[htbp]
\centering
\caption{Meta-level prompting}
\label{table:meta_prompts}
\scalebox{0.851}{
\begin{tabular}{p{1.14\linewidth}}
\toprule
\textbf{Self-Refine}~\cite{DBLP:conf/nips/MadaanTGHGW0DPY23_selfrefine}:  
After generating $t_{a1}$, the model is asked to review its own answer using  
$t_{u2} =$ {\em ``Thank you! As an expert, what do you think about the above answer? Please provide feedback so that more accurate predictions can be made in the future.''}. \\
\midrule

\textbf{Self-Consistency}~\cite{DBLP:conf/iclr/0002WSLCNCZ23_selfconsistency}:  
The prompt $t_{u1}$ is executed multiple times to produce $\{t_{a1,1}, t_{a1,2}, \ldots\}$.  
These answers are then summarized using  
$t_{u2} =$ {\em ``We asked the monitors and got the following answers: \# Collected answers: $\{t_{a1,1}, t_{a1,2}, \ldots\}$''}. \\
\bottomrule
\end{tabular}
}
\end{table}

Since these prompts require additional steps or repeated inference, their inference cost can be several times higher. To keep the comparison practical, we reduce the number of users from 100 to 10 per dataset and user type, resulting in 160 users in total\footnote{This setting balances computational cost and statistical reliability. Clear differences were already observed with this sample size in Table~\ref{table:combo_metric_improvement}.}.

\begin{table}[htbp]
\caption{nDCG@$3$ across datasets, where \textbf{SR} and \textbf{SC} refer to Self-Refine and Self-Consistency.}
\label{table:combo_metric_improvement}
\centering
\scalebox{0.83}{
\begin{tabular}{c|ccccc}
\toprule
 & 4.1-mini & llama3.3 & 4o-mini & phi-4 & nova-lite \\
\midrule
ReAct & \cellcolor{pshigh}$0.547^{*}$ & \cellcolor{pshigh}$0.533^{*}$ & \cellcolor{nhigh}$0.407$ & \cellcolor{nshigh}$0.407$ & \cellcolor{nshigh}$0.367$ \\
SC (ReAct) & \cellcolor{pshigh}$0.535^{*}$ & \cellcolor{pshigh}$0.537^{*}$ & \cellcolor{nshigh}$0.389$ & \cellcolor{nshigh}$0.402$ & \cellcolor{phigh}$0.441$ \\
SC (Step-Back) & \cellcolor{pshigh}$0.543^{*}$ & \cellcolor{pshigh}$0.518^{*}$ & \cellcolor{plow}$0.454$ & \cellcolor{nhigh}$0.427$ & \cellcolor{nshigh}$0.369$ \\
Step-Back & \cellcolor{phigh}$0.516$ & \cellcolor{pshigh}$0.543^{*}$ & \cellcolor{pslow}$0.444$ & \cellcolor{nlow}$0.439$ & \cellcolor{nshigh}$0.359$ \\
SC (Rephrase) & \cellcolor{phigh}$0.522^{*}$ & \cellcolor{pshigh}$0.536^{*}$ & \cellcolor{pmiddle}$0.468$ & $0.460$ & \cellcolor{nlow}$0.397$ \\
SR (ReAct) & \cellcolor{pshigh}$0.540^{*}$ & \cellcolor{pshigh}$0.516^{*}$ & \cellcolor{nshigh}$0.378$ & \cellcolor{nshigh}$0.369^{\bigtriangledown}$ & \cellcolor{nslow}$0.407$ \\
SR (Rephrase) & \cellcolor{phigh}$0.514$ & \cellcolor{pshigh}$0.533^{*}$ & \cellcolor{pmiddle}$0.462$ & \cellcolor{pslow}$0.471$ & \cellcolor{pshigh}$\underline{0.455}$ \\
SR (Step-Back) & \cellcolor{phigh}$0.520$ & \cellcolor{pshigh}$0.527^{*}$ & \cellcolor{plow}$0.458$ & \cellcolor{nshigh}$0.392^{\bigtriangledown}$ & \cellcolor{nmiddle}$0.390$ \\
Rephrase$\rightarrow$ReAct & \cellcolor{pshigh}$0.534^{*}$ & \cellcolor{phigh}$0.506$ & \cellcolor{nshigh}$0.376$ & \cellcolor{pslow}$0.472$ & \cellcolor{nslow}$0.402$ \\
Rephrase$\rightarrow$Step-Back & \cellcolor{phigh}$0.514$ & \cellcolor{phigh}$0.501$ & \cellcolor{pmiddle}$0.470$ & \cellcolor{plow}$0.481$ & \cellcolor{nshigh}$0.352$ \\
Baseline & $0.475$ & $0.465$ & $0.440$ & $0.460$ & $0.412$ \\
\bottomrule
\end{tabular}
}
\end{table}

According to Table~\ref{table:combo_metric_improvement}, few prompt combinations or meta-level methods outperformed the original prompts. In \texttt{gpt-4.1-mini}, applying \textbf{Self-Consistency} to \textbf{Step-Back} slightly improved accuracy, but similar effects were not observed in other cases. In \texttt{amazon-nova-lite}, applying \textbf{Self-Refine} to \textbf{Rephrase} led to a large improvement in accuracy. However, the variance of the outputs was high, and the $p$-value was not significant. Furthermore, the resulting accuracy did not exceed that of other LLMs. These results suggest that in cost-efficient LLMs such as \texttt{gpt-4.1-mini} and \texttt{llama3.3-70b}, \textbf{Rephrase}, \textbf{ReAct}, and \textbf{Step-Back} are already effective without additional techniques. Since additional instructions increased inference cost but did not meaningfully improve accuracy, this conclusion seems reasonable.

In summary, using the original forms of \textbf{Rephrase}, \textbf{ReAct}, and \textbf{Step-Back} remains the most practical choice, as they are simple, effective, and cost-efficient.

\subsection{Evaluation Using High-Performance LLMs}
\label{subsec:high_performance_model}

In the previous experiments, we used cost-efficient LLMs such as \texttt{gpt-4.1-mini} to compare all prompts. As another way to improve recommendation accuracy, we consider using more powerful LLMs. In this section, we examine whether the prompts \textbf{Rephrase}, \textbf{ReAct}, and \textbf{Step-Back}, which showed strong performance, are still effective with stronger models. For reasoning models, we evaluated only the \textbf{Baseline}, since their internal thinking processes may conflict with custom instructions, and we observed that \texttt{o3-mini} actually did.

Table~\ref{table:high_model_score_cost} shows the recommendation accuracy and inference cost for each model and prompt under the same settings as in Section~\ref{table:combo_metric_improvement}. Figure~\ref{fig:cost_map_middle} presents a two-dimensional plot of these results.

\begin{table}[htbp]
\caption{nDCG@$3$ and inference cost (USD) for each LLM. \texttt{claude-3.7-sonnet (T)} refers to the version with thinking mode enabled and is treated as a reasoning model.}
\label{table:high_model_score_cost}
\centering
\scalebox{0.76}{
\begin{tabular}{c|cccc}
\toprule
 & Baseline & Rephrase & Step-Back & ReAct \\
\midrule
llama3.3-70b & $0.465$ / 0.78 & \cellcolor{pshigh}$\underline{0.548^{*}}$ / 1.74 & \cellcolor{pshigh}$0.543^{*}$ / 1.69 & \cellcolor{pshigh}$0.533^{*}$ / 1.78 \\
gpt-4.1-mini & $0.475$ / 0.43 & \cellcolor{pshigh}$\underline{0.550^{*}}$ / 1.15 & \cellcolor{phigh}$0.516$ / 1.08 & \cellcolor{pshigh}$0.547^{*}$ / 1.18 \\
\midrule
claude-3.5-haiku & $\underline{0.516}$ / 1.03 & \cellcolor{nslow}$0.501$ / 2.28 & \cellcolor{nslow}$0.506$ / 2.30 & $0.516$ / 2.41 \\
gpt-4.1 & $0.526$ / 2.13 & \cellcolor{nslow}$0.515$ / 4.89 & $0.527$ / 4.87 & \cellcolor{pmiddle}$\underline{0.561}$ / 5.76 \\
claude-3.7-sonnet & $0.621$ / 3.85 & $0.621$ / 9.23 & \cellcolor{pslow}$\underline{0.628}$ / 8.71 & \cellcolor{nslow}$0.609$ / 10.69 \\
\midrule
o3-mini & $\underline{0.552}$ / 2.79 & \cellcolor{nslow}$0.537$ / 5.00 & \cellcolor{nslow}$0.542$ / 4.82 & (failed) \\
o4-mini & $\underline{0.537}$ / 2.09 & - & - & - \\
o3 & $\underline{0.612}$ / 11.36 & - & - & - \\
claude-3.7-sonnet (T) & $\underline{0.555}$ / 7.72 & - & - & - \\
\bottomrule
\end{tabular}
}
\end{table}

\begin{figure}[htbp]
\centering
\includegraphics[width=0.47\textwidth]{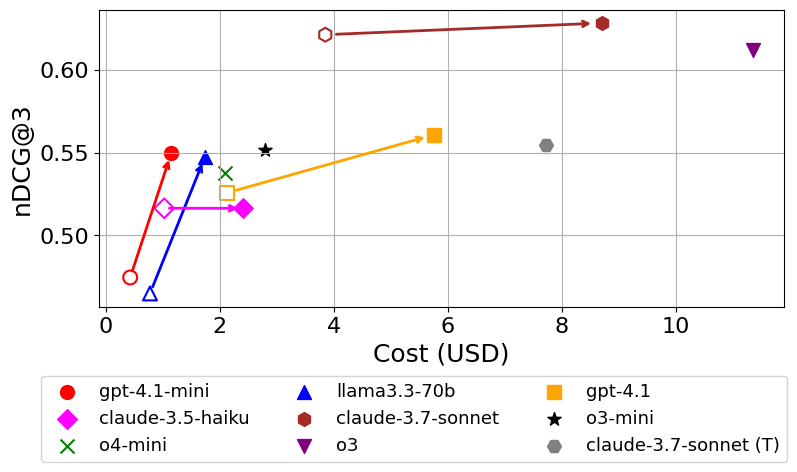}
\caption{nDCG@$3$ and inference cost. For non-reasoning models, hollow and filled markers represent the \textbf{Baseline} and the highest-scoring prompt in Table~\ref{table:high_model_score_cost}, respectively.}
\label{fig:cost_map_high}
\end{figure}

Table~\ref{table:high_model_score_cost} shows that the highest accuracy was achieved by the combination of \texttt{claude-3.7-sonnet} and \textbf{Step-Back}, which resulted in an nDCG@$3$ score of 0.628. However, other prompts using the same LLM, including \textbf{Baseline}, also had scores around 0.62, suggesting no clear advantage. The next highest score was 0.612 for \texttt{o3} with \textbf{Baseline}. Several other combinations had scores near 0.55, such as \texttt{gpt-4.1} with \textbf{ReAct}, \texttt{claude-3.7-sonnet (T)} with \textbf{Baseline}, and \texttt{o3-mini} with \textbf{Baseline}.

\underline{Accuracy decreasing from overthinking.}  
Using the thinking mode (T) in \texttt{claude-3.7-sonnet} did not improve accuracy and even resulted in lower performance than the \textbf{Baseline}. In \texttt{o3-mini}, \textbf{Rephrase} and \textbf{Step-Back} also failed to improve performance, and every attempt to use \textbf{ReAct} was blocked due to policy violations, even after five retries. A similar issue has been reported in the OpenAI developer community\footnote{\url{https://community.openai.com/t/your-prompt-was-flagged-as-potentially-violating-our-usage-policy/1060163/27}}. Such failures may be caused by internal instructions that restrict the use of certain prompts. These observations suggest that when reasoning steps become complicated, the LLM may become confused, which is referred to as the overthinking problem~\cite{DBLP:journals/corr/abs-2503-16419_stop_overthinking,DBLP:conf/iclr/HalawiDS24_overthinking}. To avoid these, we recommend checking the behavior of prompts in advance before deploying them in recommendation tasks.

\underline{Best Option for Accuracy}
Among the LLM and prompt combinations with nDCG@3 scores above 0.6, \texttt{claude-3.7-sonnet} with \textbf{Step-Back} achieved the highest accuracy. However, the difference compared to \textbf{Baseline} was minimal. \texttt{o3} also reached a comparable level of accuracy, but in terms of cost, using \texttt{claude-3.7-sonnet} with \textbf{Baseline} was several times cheaper. Considering the risk that unnecessary modifications may lead to reduce accuracy, we conclude that this simple configuration is the most reasonable choice.

\underline{Best option for cost efficiency.} For second-tier accuracy around 0.55, using \texttt{gpt-4.1-mini} or \texttt{llama3.3-70b} with one of \textbf{Rephrase}, \textbf{ReAct}, or \textbf{Step-Back} was cost-efficient. For example, \texttt{gpt-4.1-mini} with \textbf{Rephrase} achieved accuracy comparable to \texttt{claude-3.7-sonnet (T)} with \textbf{Baseline}, but at only one-fifth of the cost.

\section{Practical Guideline}
This section explains a practical guideline to apply the findings of this study to newly developed or reader-targeted LLMs, prompt methods, or datasets. As a starting point, we suggest testing four prompts that showed strong results in our experiments: \textbf{Baseline}, \textbf{Rephrase}, \textbf{Step‑Back}, and \textbf{ReAct}. At the same time, readers should record both ranking metrics such as nDCG@$k$ and Hit@$k$, and the inference cost. After collecting the results, it is important to check the inference logs. In particular, readers should look for incomplete rankings or responses that do not follow the instruction, as described in Section~\ref{subsec:error_analysis}. By comparing the four prompts, readers can select the one that fits their needs, whether they focus on accuracy, cost, or a balance between the two. We hope this guideline supports future work in LLM-based recommender systems.

\section{Conclusion}
This study examined the effect of prompt design on personalized recommendation using LLMs. We compared 23 prompts across 8 datasets and 12 LLMs in a systematic evaluation. We also tested combinations of prompts and meta-level techniques.

We found that prompts that restructure user context or guide the model through structured reasoning, such as \textbf{Rephrase}, \textbf{Step-Back}, and \textbf{ReAct}, were effective with cost-efficient LLMs like \texttt{gpt-4.1-mini} and \texttt{llama3.3-70b}. In contrast, general-purpose prompting techniques often used in NLP, such as \textbf{RolePlay}, \textbf{Emotion}, and \textbf{Step-by-Step}, did not always improve recommendation accuracy and reduced performance in some cases. For high-performance models like \texttt{claude-3.7-sonnet}, the simple \textbf{Baseline} prompt achieved high accuracy without increasing inference cost. These results provide practical guidelines for designing prompts and selecting LLMs for real-world recommendation tasks.

\newpage
\bibliographystyle{ACM-Reference-Format}
\bibliography{reference}

\clearpage

\appendix

\section{Detailed Prompts}
We show the prompts used in the main paper together with the intermediate output text from \texttt{gpt-4o-mini}.  We replace long texts where the LLM examines many items and its inner reasoning steps with the blue marker ``(omitted)''.

\begin{itemize}
    \item \textbf{Baseline}: Figure~\ref{list:baseline}
    \item \textbf{Rephrase}: Figure~\ref{list:rephrase}
    \item \textbf{ReAct}: Figure~\ref{list:react}
    \item \textbf{Step-Back}: Figure~\ref{list:step-back}
    \item \textbf{Step-by-Step}: Figure~\ref{list:step-by-step}
    \item \textbf{Pretend-User}: Figure~\ref{list:pretend}
    \item \textbf{Summarize-Item}: Figure~\ref{list:summarize-item}
    \item \textbf{Rephrase} $\rightarrow$ \textbf{ReAct}: Figure~\ref{list:Rephrase_ReAct}
    \item \textbf{SelfRefine} (\textbf{Step-Back}): Figure~\ref{list:selfrefine_step-back}
\end{itemize}

\section{Accuracies in Section 4.1}
Table~7 in the main paper showed the nDCG@3 accuracy for all datasets using \texttt{gpt-4.1-mini}. In this section, we show accuracy tables for the other LLMs.
\begin{itemize}
    \item \texttt{llama3.3-70b}: Table~\ref{table:llama3.3-70b}
    \item \texttt{gpt-4o-mini}: Table~\ref{table:gpt-4o-mini} 
    \item \texttt{phi4}: Table~\ref{table:phi4}
    \item \texttt{amazon-nova-lite}: Table~\ref{table:amazon-nova-lite}
\end{itemize}

We also present the prompts and LLMs used with each dataset.

\begin{itemize}
    \item \textbf{Music}: Table~\ref{table:music}
    \item \textbf{Movie}: Table~\ref{table:movie}
    \item \textbf{Groceries}: Table~\ref{table:groceries}
    \item \textbf{Clothes}: Table~\ref{table:clothes}
    \item \textbf{Books}: Table~\ref{table:books}
    \item \textbf{Yelp}: Table~\ref{table:yelp}
    \item \textbf{News}: Table~\ref{table:news}
    \item \textbf{Food}: Table~\ref{table:food}
    \item \textbf{Light}: Table~\ref{table:light}
    \item \textbf{Heavy}: Table~\ref{table:heavy}
\end{itemize}

Table~\ref{table:hit3} and Table~\ref{table:hit3_high} show the Hit@3 results corresponding to Table~7 and Table~13 in the main paper.

\section{Error Analysis in Section 4.3}

Table~10 in the main paper shows the percentage of cases where the LLMs generated three or fewer ranked items, as well as the percentage of failed cases. Table~\ref{table:partial_ranking_1_5} shows the percentage of cases where the LLMs generated five or fewer and nine or fewer items.

Table~\ref{table:token_stat_baseline} and Table~\ref{table:token_stat_itemsummary} show the maximum, minimum, average, and standard deviation of token counts for the \textbf{Baseline} and \textbf{Summarize-Item} prompts across all datasets.

Figure~\ref{list:fail_ReAct} shows the internal reasoning process in a failed case using the \textbf{ReAct} prompt with \texttt{gpt-4o-mini}. In this case, the LLM responded with ``{\em Since none of the candidate products align with the user's preferences}''.

\section{Inference Cost in the main paper}
Table~\ref{table:cost_section4}, Table~\ref{table:cost_section51} and Table~\ref{table:cost_section52} summarize the inference costs in Section 4, 5.1, and 5.2 of the main paper.

\clearpage

\begin{figure*}[htbp]
\centering
\begin{lstlisting}[frame=single]
(*@{$t_{\textrm{sys}}$@*) = You are an AI assistant that helps people find information.
--------------------------------------------------------------------------------------------------------------------------------
(*@{$t_{u1}$@*) = # Requirements:
you must rank candidate products that will be provided below to the target user for recommendation.

# Observation:
{1: {'title': 'Bourne Ultimatum', 'categories': 'Studio Specials, Universal Studios Home Entertainment, All Universal Studios Titles', 'description': '', 'rating': 5.0, 'review': 'awesome movie'}, 
(*@\textcolor{blue}{(omitted)}@*)
5: {'title': 'The Bourne Legacy', 'categories': 'Fully Loaded DVDs, DTS', 'description': 'The Bourne Legacy takes the action-packed Bourne series to an explosive new level. On the verge of having their conspiracy exposed, members of the government\'s intelligence community will stop at nothing to erase all evidence of their top secret programs - even the agents involved. Aaron Cross (Jeremy Renner) must use his genetically-engineered skills to survive the ultimate game of cat-and-mouse and finish what Jason Bourne started. Also starring Academy Award winner Rachel Weisz and Academy Award nominee Edward Norton, critics are calling this a "thrilling, edge-of-your-seat heart-pounder" (Meg Porter Berns, WSVN-TV (FOX), Miami).', 'rating': 5.0, 'review': 'awesome movie'}}

Based on the above user information, please rank the candidate products that align closely with the user's preferences. If item IDs [101, 102, 103] are sorted as first 102, second 103, third 101 in order, present your response in the format below: [3,1,2]. In this example, the number of candidate products was 3, but next, 10 products will be provided, so you SHOULD sort 10 products. 

# Candidates products:
1: {'title': 'Chicago', 'categories': 'Studio Specials, Miramax Home Entertainment, All Titles', 'description': 'Earning 13 Academy Award nominations including Best Picture, Best Director, Best Actress, and Best Supporting Actress, Chicago stars Catherine Zeta-Jones, Renee Zellweger and Richard Gere. Directed by Rob Marshall, it&#8217;s a dazzling spectacle that was cheered by both audiences and critics nationwide. PG-13. 107 minutes. DVD.'}, 
(*@\textcolor{blue}{(omitted)}@*)
10: {'title': 'Love Streams', 'categories': 'Criterion Collection, All', 'description': 'A troubled divorcee and a moody writer are apart and then together, sharing a special relation.'}}

I here repeat candidate instruction. Please rank 10 candidate product IDs as follows: If product IDs [101, 102, 103] are sorted as first 102, second 103, third 101 in order, present your response in the format below: [3,1,2].
        
# Additional constraints:
- You ONLY rank the given Candidate product IDs. 
- Do not explain the reason and include any other words. 
\end{lstlisting}
\caption{Baseline}
\label{list:baseline}
\end{figure*}

\begin{figure*}[htbp]
\centering
\begin{lstlisting}[frame=single]
(*@{$t_{\textrm{sys}}$@*) = (*@\textcolor{blue}{(default\_system\_inst)}@*)
--------------------------------------------------------------------------------------------------------------------------------
(*@{$t_{u1}$@*) = (*@\textcolor{blue}{(user\_info + candidate\_info)}@*)

Our final goal is to provide an answer to the following problem. Before tackling this issue, rephrase and expand it to help you do better answering. Maintain all information in the original question.
--------------------------------------------------------------------------------------------------------------------------------
(*@{$t_{a1}$@*) = The task at hand is to evaluate and rank a list of candidate products based on their alignment with the preferences of a target user. The user has previously shown a strong interest in action-packed films, particularly those in the Bourne series, as well as comedies with high ratings. The goal is to identify which of the new candidate products best match these interests and rank them accordingly.

The candidate products provided include a mix of genres, including action, comedy, and drama. Each product has a title, category, and description, which will help determine how closely they align with the user's preferences. 

The ranking should reflect the following considerations:
1. The presence of action elements, particularly in the context of the Bourne series.
2. High ratings and positive reviews, as the user has previously favored products with a perfect rating of 5.0.
3. The relevance of the genre to the user's interests, with a focus on action and comedy.

After analyzing the candidate products, I will provide a ranked list indicating the order of preference for the user. 

Now, let's proceed to rank the candidate products based on these criteria. 

Here are the candidate products to be ranked:
1. Seven Year Itch VHS (Comedy)
(*@\textcolor{blue}{(omitted)}@*)
10. Love Streams (Drama)

Based on the analysis of these products, the final ranking will be provided in the specified format. 

Now, let's rank the products. 

### Ranked List:
1. **The Bourne Identity** (Action) - Strong alignment with the user's interest in the Bourne series.
(*@\textcolor{blue}{(omitted)}@*)
10. **Cooking French, Vol. 2** (Special Interests) - Least relevant to the user's interests.

### Final Output:
[4, 9, 8, 6, 1, 3, 2, 7, 10, 5]
--------------------------------------------------------------------------------------------------------------------------------
(*@{$t_{u2}$@*) = (*@\textcolor{blue}{(final\_inst)}@*)
\end{lstlisting}
\caption{Rephrase}
\label{list:rephrase}
\end{figure*}

\begin{figure*}[htbp]
\centering
\begin{lstlisting}[frame=single]
(*@{$t_{\textrm{sys}}$@*) = (*@\textcolor{blue}{(default\_system\_inst)}@*)
--------------------------------------------------------------------------------------------------------------------------------
(*@{$t_{u1}$@*) = (*@\textcolor{blue}{(user\_info + candidate\_info)}@*)

Our final goal is to provide an answer to the following problem. Before tackling this issue, please follow this format to proceed step by step with *Observation*, *Thought*, and *Action*:
- Observation: Observe the user's history and preferences.
- Thought: Infer the user's tastes or tendencies from the observation.
- Action: Choose one candidate item and examine its characteristics.
- Observation: Observe the characteristics of that item.
- Thought: Consider whether the item matches the user's preferences. 
(Repeat for multiple items if necessary)
Finally, provide your *Answer*.
--------------------------------------------------------------------------------------------------------------------------------
(*@{$t_{a1}$@*) = ### Observation:
The user has shown a strong preference for action-packed movies, particularly those in the Bourne series, as well as high-rated comedies like "Tyler Perry's A Madea Christmas." All items in the user's history have a perfect rating of 5.0 and are well-reviewed, indicating a preference for quality content.

### Thought:
The user likely enjoys thrilling action films and comedies with strong narratives and engaging characters. Given this, I will prioritize candidate products that are action-oriented or highly rated comedies.

### Action:
I will examine the candidate items one by one to determine their relevance to the user's preferences.

1. **Seven Year Itch VHS**: Comedy, but lacks a strong description or notable acclaim.
(*@\textcolor{blue}{(omitted)}@*)
10. **Love Streams**: Drama, not aligned with the user's preferences.

### Observation:
- **The Bourne Identity**: Strongly matches user preferences as it is an action film from a series they enjoy.
- **Underworld: Awakening**: Action-oriented, but may not be as appealing as the Bourne series.
- **Jack Reacher Combo**: Another action film that fits the user's taste for thrillers.

### Thought:
The top candidates based on the user's preferences are "The Bourne Identity," "Jack Reacher Combo," and "Underworld: Awakening." "The Bourne Identity" is the most aligned with the user's established preferences, followed by "Jack Reacher Combo" and then "Underworld: Awakening."

### Answer:
[4, 9, 8]
--------------------------------------------------------------------------------------------------------------------------------
(*@{$t_{u2}$@*) = (*@\textcolor{blue}{(final\_inst)}@*)
\end{lstlisting}
\caption{ReAct}
\label{list:react}
\end{figure*}

\begin{figure*}[htbp]
\centering
\begin{lstlisting}[frame=single]
(*@{$t_{\textrm{sys}}$@*) = (*@\textcolor{blue}{(default\_system\_inst)}@*)
--------------------------------------------------------------------------------------------------------------------------------
(*@{$t_{u1}$@*) = (*@\textcolor{blue}{(user\_info + candidate\_info)}@*)

Our final goal is to provide an answer to the following problem. Before tackling this issue, please consider the principles and theories behind this question.
--------------------------------------------------------------------------------------------------------------------------------
(*@{$t_{a1}$@*) = To rank the candidate products based on the user's preferences, we should consider the following factors:

1. **Genre and Themes**: The user has shown a preference for action and thriller movies, particularly those in the Bourne series, which are characterized by intense action and espionage themes.
2. **Ratings**: Products with higher ratings (5.0 in this case) are more favorable.
3. **Familiarity with Series**: The user has already shown interest in the Bourne series, so products related to that series should be prioritized.
4. **Overall Appeal**: Products that are critically acclaimed or have a strong following may also be ranked higher.

Now, let's analyze the candidate products:

1. **The Bourne Identity** (4) - This is directly related to the Bourne series and fits the user's established interest in action and espionage.
(*@\textcolor{blue}{(omitted)}@*)
10. **Love Streams** (10) - A drama that does not fit the user's established preferences for action and thrillers.

Based on this analysis, the ranked list of candidate products is:

[4, 8, 9, 2, 3, 6, 7, 1, 5, 10]
--------------------------------------------------------------------------------------------------------------------------------
(*@{$t_{u2}$@*) = (*@\textcolor{blue}{(final\_inst)}@*)
\end{lstlisting}
\caption{Step-Back}
\label{list:step-back}
\end{figure*}

\begin{figure*}[htbp]
\centering
\begin{lstlisting}[frame=single]
(*@{$t_{\textrm{sys}}$@*) = (*@\textcolor{blue}{(default\_system\_inst)}@*)
--------------------------------------------------------------------------------------------------------------------------------
(*@{$t_{u1}$@*) = (*@\textcolor{blue}{(user\_info + candidate\_info)}@*)
--------------------------------------------------------------------------------------------------------------------------------
(*@{$t_{a1}$@*) = Let's think step by step.
--------------------------------------------------------------------------------------------------------------------------------
(*@{$t_{u2}$@*) = OK! Go ahead.
--------------------------------------------------------------------------------------------------------------------------------
(*@{$t_{a2}$@*) = To rank the candidate products, we will consider the following factors based on the user's previous preferences:

1. **Genre**: The user seems to prefer action and thriller movies, particularly those related to the Bourne series.
2. **Rating**: Higher-rated products are more likely to be recommended.
3. **Description**: Products with engaging descriptions that align with the user's interests will be prioritized.

Now, let's analyze the candidate products:

1. **The Bourne Identity** (Product 4): This is directly related to the Bourne series, which the user has shown a strong preference for. It has a compelling description and fits the action/thriller genre.
(*@\textcolor{blue}{(omitted)}@*)
10. **Love Streams** (Product 10): This is a drama that may not resonate with the user's action/thriller interests.

Based on this analysis, the ranked list of candidate products is:

[4, 8, 9, 2, 3, 6, 1, 7, 5, 10]
--------------------------------------------------------------------------------------------------------------------------------
(*@{$t_{u3}$@*) = (*@\textcolor{blue}{(final\_inst)}@*)
\end{lstlisting}
\caption{Step-by-Step}
\label{list:step-by-step}
\end{figure*}

\begin{figure*}[htbp]
\centering
\begin{lstlisting}[frame=single]
(*@{$t_{\textrm{sys}}$@*) = You are an AI assistant that pretends to be a person who has interacted with the following items.
# Logs: (*@\textcolor{blue}{(user\_item\_history)}@*)
--------------------------------------------------------------------------------------------------------------------------------
(*@{$t_{u1}$@*) = Please rank the candidate products that align closely with your preferences. If item IDs [101, 102, 103] are sorted as first 102, second 103, third 101 in order, present your response in the format below: [3,1,2]. In this example, the number of candidate products was 3, but next, 10 products will be provided, so you SHOULD sort 10 products. 

# Candidates products:(*@\textcolor{blue}{(candidate\_items)}@*)

(*@\textcolor{blue}{(final\_inst)}@*)
\end{lstlisting}
\caption{Pretend-User}
\label{list:pretend}
\end{figure*}

\begin{figure*}[htbp]
\centering
\begin{lstlisting}[frame=single]
(*@{$t_{\textrm{sys}}$@*) = (*@\textcolor{blue}{(default\_system\_inst)}@*)
--------------------------------------------------------------------------------------------------------------------------------
(*@{$t_{u1}$@*) = # Requirements:
you must rank candidate products that will be provided below to the target user for recommendation.

# Observation:
{1: {'summary': '**Item Summary: "Bourne Ultimatum"**\n\n- **Title:** Bourne Ultimatum\n- **Categories:** This item falls under several categories, including Studio Specials, Universal Studios Home Entertainment, and All Universal Studios Titles. This classification highlights its association with a prominent studio known for quality entertainment.\n- **Description:** Currently, there is no detailed description provided for the item.\n\n"Bourne Ultimatum" is a notable film that is part of the acclaimed Bourne series, recognized for its thrilling action and intricate plot. As part of Universal Studios\' collection, it reflects the studio\'s commitment to delivering high-quality cinematic experiences.', 'rating': 5.0, 'review': 'awesome movie'}, 
(*@\textcolor{blue}{(omitted)}@*)
5: {'summary': '**Title:** The Bourne Legacy  \n**Categories:** Fully Loaded DVDs, DTS  \n**Description:** "The Bourne Legacy" elevates the action-packed Bourne series as it follows Aaron Cross (Jeremy Renner), a genetically-engineered agent, who must navigate a dangerous game of cat-and-mouse to survive after the government\'s intelligence community seeks to eliminate all evidence of their covert programs. With a star-studded cast including Academy Award winners Rachel Weisz and Edward Norton, the film is praised as a "thrilling, edge-of-your-seat heart-pounder" by critics. This installment promises explosive action and suspense, making it a must-watch for fans of the franchise.', 'rating': 5.0, 'review': 'awesome movie'}}

(*@\textcolor{blue}{(candidate\_info)}@*)
(*@\textcolor{blue}{(final\_inst)}@*)
\end{lstlisting}
\caption{Summarize-Item}
\label{list:summarize-item}
\end{figure*}

\begin{figure*}[htbp]
\centering
\begin{lstlisting}[frame=single]
(*@{$t_{\textrm{sys}}$@*) = (*@\textcolor{blue}{(default\_system\_inst)}@*)
--------------------------------------------------------------------------------------------------------------------------------
(*@{$t_{u1}$@*) = (*@\textcolor{blue}{(user\_info + candidate\_info)}@*)

Our final goal is to provide an answer to the ranking item problem. Before tackling this issue, rephrase and expand it to help you do better answering. Maintain all information in the original question.
--------------------------------------------------------------------------------------------------------------------------------
(*@{$t_{a1}$@*) = # Task Overview:
You are tasked with ranking a set of candidate products based on their alignment with user preferences. The user has previously shown a strong interest in action-packed films, particularly those in the Bourne series, as well as comedies that are well-reviewed. The goal is to recommend the top products that best match these interests.

# User Preferences:
The user has expressed a preference for movies with high ratings and engaging content. The previous products they liked include:
1. **Bourne Ultimatum** - Action-packed with a perfect rating of 5.0.
(*@\textcolor{blue}{(omitted)}@*)
5. **The Bourne Legacy** - A thrilling addition to the Bourne franchise, rated 5.0.

# Candidate Products:
You will evaluate the following 10 candidate products to determine which ones best align with the user's preferences for high-rated action and comedy films:

1. **Seven Year Itch VHS** - Comedy genre.
(*@\textcolor{blue}{(omitted)}@*)
10. **Love Streams** - A drama film.

# Objective:
Your task is to rank these 10 candidate products based on how closely they align with the user's preferences for action and comedy films, particularly those that are highly rated. The ranking should be presented in the format of item IDs sorted by their relevance to the user's interests.

Please provide the ranked list of candidate products.
--------------------------------------------------------------------------------------------------------------------------------
(*@{$t_{u2}$@*) = Our final goal is to provide an answer to the ranking item problem. Before tackling this issue, please follow this format to proceed step by step with *Observation*, *Thought*, and *Action*:
- Observation: Observe the user's history and preferences.
- Thought: Infer the user's tastes or tendencies from the observation.
- Action: Choose one candidate item and examine its characteristics.
- Observation: Observe the characteristics of that item.
- Thought: Consider whether the item matches the user's preferences. 
(Repeat for multiple items if necessary)
Finally, provide your *Answer*.
--------------------------------------------------------------------------------------------------------------------------------
(*@{$t_{a2}$@*) = ### Step-by-Step Process

**Observation:** The user has a strong preference for action-packed films, particularly those in the Bourne series, as well as comedies that are well-reviewed. All previously liked products have a perfect rating of 5.0.

**Thought:** The user enjoys high-quality films that are either thrilling or humorous. The Bourne series is a clear favorite, indicating a strong preference for action and espionage themes. Comedies that are well-received also appeal to the user.

---
(*@\textcolor{blue}{(omitted)}@*)
---
(*@\textcolor{blue}{(omitted)}@*)
---
(*@\textcolor{blue}{(omitted)}@*)
---
(*@\textcolor{blue}{(omitted)}@*)
---
(*@\textcolor{blue}{(omitted)}@*)
---

**Final Evaluation of All Candidates:**
1. **The Bourne Identity** - Strong match for action preference.
(*@\textcolor{blue}{(omitted)}@*)
5. **Chicago** - Critically acclaimed but not action or comedy.

### Answer:
Based on the evaluations, the ranked list of candidate products is: **[4, 9, 8, 3, 2]**.
--------------------------------------------------------------------------------------------------------------------------------
(*@{$t_{u3}$@*) = (*@\textcolor{blue}{(final\_inst)}@*)
\end{lstlisting}
\caption{Rephrase $\rightarrow$ ReAct}
\label{list:Rephrase_ReAct}
\end{figure*}

\begin{figure*}[htbp]
\centering
\begin{lstlisting}[frame=single]
(*@{$t_{\textrm{sys}}$@*) = (*@\textcolor{blue}{(default\_system\_inst)}@*)
--------------------------------------------------------------------------------------------------------------------------------
(*@{$t_{u1}$@*) = (*@\textcolor{blue}{(user\_info + candidate\_info)}@*)

Our final goal is to provide an answer to the ranking item problem. Before tackling this issue, please consider the principles and theories behind the first question.
--------------------------------------------------------------------------------------------------------------------------------
(*@{$t_{a1}$@*) = To rank the candidate products based on the user's preferences, we will focus on the following criteria:

1. **Genre and Themes**: The user has shown a preference for action and thriller movies, particularly those in the Bourne series, which are characterized by intense action and espionage themes.
2. **Ratings**: Higher-rated products (5.0 ratings) are prioritized.
3. **Relevance**: Products that are part of the same franchise or have similar themes to the Bourne series will be ranked higher.

Given these criteria, let's analyze the candidate products:

1. **The Bourne Identity** (Product 4): This is directly related to the Bourne series, which the user has shown a strong preference for. It has a similar action and espionage theme.
(*@\textcolor{blue}{(omitted)}@*)
10. **Love Streams** (Product 10): This is a drama and does not fit the user's action-oriented preferences.

Based on this analysis, the ranked list of candidate products is as follows:

[4, 8, 9, 2, 3, 6, 7, 1, 5, 10]
--------------------------------------------------------------------------------------------------------------------------------
(*@{$t_{u2}$@*) = Thank you! As an expert, what do you think about the above answers? Please provide feedback so that more accurate predictions can be made in the future.
--------------------------------------------------------------------------------------------------------------------------------
(*@{$t_{a2}$@*) = The ranking provided is a solid attempt to align the candidate products with the user's preferences based on the information available. However, there are several areas for improvement to enhance the accuracy of future predictions:

1. **User Preference Analysis**: While the analysis focused on action and thriller genres, it could benefit from a deeper understanding of the user's overall preferences. For instance, if the user enjoys a mix of genres, including comedy or drama, those products could be ranked higher.

(*@\textcolor{blue}{(omitted)}@*)

7. **Feedback Loop**: Implementing a feedback mechanism where users can indicate their satisfaction with recommendations could help refine future predictions. This would allow for continuous learning and improvement based on user responses.

By incorporating these considerations, future rankings can become more nuanced and tailored to the user's preferences, leading to more accurate and satisfying recommendations.
--------------------------------------------------------------------------------------------------------------------------------
(*@{$t_{u3}$@*) = (*@\textcolor{blue}{(final\_inst)}@*)
\end{lstlisting}
\caption{SelfRefine (Step-Back)}
\label{list:selfrefine_step-back}
\end{figure*}

\clearpage

\begin{table*}[htbp]
\caption{llama3.3-70b}
\label{table:llama3.3-70b}
\centering
\scalebox{1}{
\begin{tabular}{lllllllllll}
\toprule
 & Music & Movie & Groceries & Clothes & Book & Yelp & News & Food & Heavy & Light \\
\midrule
Baseline & $0.605$ & $0.506$ & $0.478$ & $0.408$ & $0.574$ & $0.475$ & $0.157$ & $0.312$ & $0.440$ & $0.439$ \\
Emotion & $0.607$ & \cellcolor{pslow}$0.514$ & \cellcolor{pslow}$0.484$ & $0.406$ & $0.575$ & \cellcolor{pslow}$0.484$ & \cellcolor{pshigh}$0.177$ & \cellcolor{nmiddle}$0.290$ & $0.439$ & \cellcolor{pslow}$0.445$ \\
Re-Reading & \cellcolor{plow}$0.631$ & \cellcolor{plow}$0.526$ & \cellcolor{pshigh}$\underline{0.554^{*}}$ & \cellcolor{phigh}$0.441$ & $0.577$ & \cellcolor{pmiddle}$0.507^{*}$ & \cellcolor{pshigh}$0.233^{*}$ & \cellcolor{nslow}$0.308$ & \cellcolor{phigh}$0.471^{*}$ & \cellcolor{phigh}$0.474^{*}$ \\
Both-Inst & $0.607$ & $0.502$ & \cellcolor{plow}$0.495$ & $0.404$ & \cellcolor{nslow}$0.561$ & \cellcolor{pmiddle}$0.507^{*}$ & \cellcolor{phigh}$0.169$ & \cellcolor{nshigh}$0.248^{\bigtriangledown}$ & $0.437$ & $0.436$ \\
Recency-Focused & \cellcolor{plow}$0.627$ & \cellcolor{plow}$0.523$ & \cellcolor{plow}$0.495$ & \cellcolor{pshigh}$\underline{0.473^{*}}$ & \cellcolor{plow}$0.601$ & \cellcolor{nslow}$0.468$ & \cellcolor{pshigh}$0.248^{*}$ & \cellcolor{plow}$0.324$ & \cellcolor{pmiddle}$0.467^{*}$ & \cellcolor{phigh}$0.473^{*}$ \\
RolePlay-User & $0.606$ & \cellcolor{pslow}$0.512$ & $0.482$ & \cellcolor{phigh}$0.446^{*}$ & \cellcolor{nslow}$0.560$ & \cellcolor{pslow}$0.481$ & \cellcolor{phigh}$0.169$ & \cellcolor{pslow}$0.315$ & \cellcolor{plow}$0.461^{*}$ & \cellcolor{nslow}$0.432$ \\
RolePlay-Expert & \cellcolor{nslow}$0.597$ & \cellcolor{pslow}$0.516$ & \cellcolor{pslow}$0.490$ & \cellcolor{nslow}$0.400$ & \cellcolor{nslow}$0.561^{\bigtriangledown}$ & \cellcolor{plow}$0.490$ & \cellcolor{nslow}$0.154$ & $0.313$ & $0.439$ & $0.441$ \\
\bottomrule
\end{tabular}
}
\end{table*}

\begin{table*}[htbp]
\caption{gpt-4o-mini}
\label{table:gpt-4o-mini}
\centering
\scalebox{1}{
\begin{tabular}{lllllllllll}
\toprule
 & Music & Movie & Groceries & Clothes & Book & Yelp & News & Food & Heavy & Light \\
\midrule
Baseline & $0.541$ & $0.502$ & $0.440$ & $0.393$ & $0.518$ & $0.501$ & $0.200$ & $0.364$ & $0.444$ & $0.421$ \\
Emotion & \cellcolor{plow}$0.561$ & \cellcolor{plow}$0.527$ & \cellcolor{pslow}$0.445$ & \cellcolor{pslow}$0.400$ & \cellcolor{plow}$0.543$ & \cellcolor{nlow}$0.479$ & \cellcolor{pslow}$0.204$ & $0.360$ & \cellcolor{pslow}$0.450$ & \cellcolor{pslow}$0.430$ \\
Re-Reading & \cellcolor{nshigh}$0.456^{\bigtriangledown}$ & \cellcolor{nlow}$0.483$ & \cellcolor{phigh}$0.471$ & \cellcolor{nlow}$0.377$ & \cellcolor{nshigh}$0.436^{\bigtriangledown}$ & \cellcolor{nshigh}$0.439^{\bigtriangledown}$ & \cellcolor{phigh}$0.219$ & \cellcolor{nhigh}$0.335$ & \cellcolor{nlow}$0.431$ & \cellcolor{nshigh}$0.374^{\bigtriangledown}$ \\
Both-Inst & \cellcolor{phigh}$0.586^{*}$ & \cellcolor{nlow}$0.484$ & $0.437$ & $0.397$ & $0.520$ & \cellcolor{nlow}$0.485$ & \cellcolor{nslow}$0.194$ & \cellcolor{nmiddle}$0.341$ & \cellcolor{nslow}$0.432^{\bigtriangledown}$ & \cellcolor{pslow}$0.429$ \\
Recency-Focused & \cellcolor{plow}$0.565$ & $0.497$ & \cellcolor{pmiddle}$0.466$ & \cellcolor{pslow}$0.398$ & \cellcolor{pslow}$0.532$ & \cellcolor{nlow}$0.480$ & \cellcolor{pshigh}$0.230$ & \cellcolor{plow}$0.375$ & \cellcolor{pslow}$0.449$ & \cellcolor{plow}$0.437$ \\
RolePlay-User & \cellcolor{phigh}$0.591^{*}$ & \cellcolor{nslow}$0.492$ & \cellcolor{nshigh}$0.358^{\bigtriangledown}$ & \cellcolor{nshigh}$0.344^{\bigtriangledown}$ & \cellcolor{pmiddle}$0.545$ & \cellcolor{nslow}$0.493$ & \cellcolor{nshigh}$0.116^{\bigtriangledown}$ & \cellcolor{nlow}$0.348$ & \cellcolor{nlow}$0.426^{\bigtriangledown}$ & \cellcolor{nmiddle}$0.396^{\bigtriangledown}$ \\
RolePlay-Expert & \cellcolor{plow}$0.566$ & $0.503$ & $0.441$ & \cellcolor{nslow}$0.383$ & \cellcolor{nslow}$0.513$ & $0.505$ & \cellcolor{nlow}$0.191$ & \cellcolor{nmiddle}$0.344$ & $0.443$ & $0.418$ \\
\bottomrule
\end{tabular}
}
\end{table*}

\begin{table*}[htbp]
\caption{phi4}
\label{table:phi4}
\centering
\scalebox{1}{
\begin{tabular}{lllllllllll}
\toprule
 & Music & Movie & Groceries & Clothes & Book & Yelp & News & Food & Heavy & Light \\
\midrule
Baseline & $0.591$ & $0.462$ & $0.464$ & $\underline{0.471}$ & $0.595$ & $0.331$ & $0.145$ & $0.278$ & $0.423$ & $0.411$ \\
Emotion & \cellcolor{nslow}$0.573$ & \cellcolor{nlow}$0.446$ & \cellcolor{plow}$0.483$ & \cellcolor{nlow}$0.451$ & \cellcolor{nslow}$0.580$ & \cellcolor{pmiddle}$0.352$ & \cellcolor{nhigh}$0.131$ & \cellcolor{pshigh}$\underline{0.319^{*}}$ & \cellcolor{pslow}$0.428$ & \cellcolor{nslow}$0.406$ \\
Re-Reading & \cellcolor{nhigh}$0.540^{\bigtriangledown}$ & \cellcolor{nslow}$0.457$ & \cellcolor{plow}$0.478$ & $0.469$ & \cellcolor{nshigh}$0.533^{\bigtriangledown}$ & \cellcolor{nhigh}$0.306$ & \cellcolor{pshigh}$0.244^{*}$ & \cellcolor{nshigh}$0.244$ & \cellcolor{pslow}$0.432$ & \cellcolor{nmiddle}$0.386^{\bigtriangledown}$ \\
Both-Inst & \cellcolor{nmiddle}$0.553^{\bigtriangledown}$ & \cellcolor{nslow}$0.455$ & $0.462$ & \cellcolor{nhigh}$0.435^{\bigtriangledown}$ & \cellcolor{nslow}$0.580$ & \cellcolor{phigh}$0.360$ & \cellcolor{nshigh}$0.083^{\bigtriangledown}$ & \cellcolor{pmiddle}$0.293$ & \cellcolor{nslow}$0.417$ & \cellcolor{nmiddle}$0.388^{\bigtriangledown}$ \\
Recency-Focused & \cellcolor{nlow}$0.562$ & \cellcolor{nslow}$0.455$ & \cellcolor{nslow}$0.457$ & \cellcolor{nhigh}$0.429$ & $\underline{0.598}$ & \cellcolor{pmiddle}$0.352$ & \cellcolor{nhigh}$0.131$ & \cellcolor{pmiddle}$0.292$ & $0.424$ & \cellcolor{nlow}$0.395$ \\
RolePlay-User & \cellcolor{nshigh}$0.529^{\bigtriangledown}$ & \cellcolor{nhigh}$0.423^{\bigtriangledown}$ & \cellcolor{nshigh}$0.395^{\bigtriangledown}$ & \cellcolor{nshigh}$0.399^{\bigtriangledown}$ & \cellcolor{nlow}$0.573$ & \cellcolor{nlow}$0.320$ & \cellcolor{nshigh}$0.042^{\bigtriangledown}$ & \cellcolor{pshigh}$0.316$ & \cellcolor{nhigh}$0.391^{\bigtriangledown}$ & \cellcolor{nshigh}$0.358^{\bigtriangledown}$ \\
\bottomrule
\end{tabular}
}
\end{table*}

\begin{table*}[htbp]
\caption{amazon-nova-lite}
\label{table:amazon-nova-lite}
\centering
\scalebox{1}{
\begin{tabular}{lllllllllll}
\toprule
 & Music & Movie & Groceries & Clothes & Book & Yelp & News & Food & Heavy & Light \\
\midrule
Baseline & $0.441$ & $0.401$ & $0.414$ & $0.367$ & $0.470$ & $\underline{0.447}$ & $0.267$ & $0.305$ & $0.360$ & $0.418$ \\
Emotion & \cellcolor{nlow}$0.425$ & \cellcolor{plow}$0.418$ & \cellcolor{nhigh}$0.382$ & \cellcolor{plow}$0.385$ & \cellcolor{nslow}$0.461$ & \cellcolor{nhigh}$0.404^{\bigtriangledown}$ & $0.265$ & \cellcolor{nshigh}$0.266$ & \cellcolor{nlow}$0.347$ & \cellcolor{nlow}$0.404$ \\
Re-Reading & \cellcolor{pshigh}$\underline{0.509^{*}}$ & \cellcolor{nmiddle}$0.381$ & \cellcolor{pslow}$\underline{0.421}$ & \cellcolor{pslow}$0.378$ & \cellcolor{pmiddle}$0.494$ & \cellcolor{nshigh}$0.389^{\bigtriangledown}$ & \cellcolor{pshigh}$0.316^{*}$ & \cellcolor{nshigh}$0.258$ & \cellcolor{pmiddle}$0.379$ & \cellcolor{nslow}$0.408$ \\
Both-Inst & \cellcolor{nlow}$0.426$ & \cellcolor{pslow}$0.413$ & \cellcolor{nhigh}$0.379$ & \cellcolor{pshigh}$0.412^{*}$ & \cellcolor{pslow}$0.475$ & \cellcolor{nlow}$0.433$ & \cellcolor{pshigh}$0.383^{*}$ & \cellcolor{nshigh}$0.249^{\bigtriangledown}$ & \cellcolor{pslow}$0.364$ & \cellcolor{pslow}$0.429$ \\
Recency-Focused & \cellcolor{phigh}$0.481$ & \cellcolor{plow}$0.415$ & $0.412$ & \cellcolor{pshigh}$\underline{0.433^{*}}$ & \cellcolor{nlow}$0.450$ & \cellcolor{nshigh}$0.402^{\bigtriangledown}$ & \cellcolor{pslow}$0.272$ & \cellcolor{nmiddle}$0.289$ & \cellcolor{nslow}$0.357$ & \cellcolor{plow}$\underline{0.432}$ \\
RolePlay-User & \cellcolor{nlow}$0.428$ & \cellcolor{phigh}$0.431$ & \cellcolor{nshigh}$0.331^{\bigtriangledown}$ & $0.369$ & \cellcolor{nhigh}$0.426$ & \cellcolor{nlow}$0.425$ & \cellcolor{pshigh}$0.439^{*}$ & \cellcolor{nhigh}$0.280$ & \cellcolor{plow}$0.372$ & \cellcolor{nslow}$0.411$ \\
\bottomrule
\end{tabular}
}
\end{table*}

\clearpage

\begin{table*}[htbp]
\caption{Music}
\label{table:music}
\centering
\scalebox{1}{
\begin{tabular}{llllll}
\toprule
 & gpt-4.1-mini & llama3.3-70b & gpt-4o-mini & phi4 & amazon-nova-lite \\
\midrule
Baseline & $0.610$ & $0.605$ & $0.541$ & $0.591$ & $0.441$ \\
Emotion & $0.608$ & $0.607$ & \cellcolor{plow}$0.561$ & \cellcolor{nslow}$0.573$ & \cellcolor{nlow}$0.425$ \\
Re-Reading & \cellcolor{nhigh}$0.567^{\bigtriangledown}$ & \cellcolor{plow}$0.631$ & \cellcolor{nshigh}$0.456^{\bigtriangledown}$ & \cellcolor{nhigh}$0.540^{\bigtriangledown}$ & \cellcolor{pshigh}$\underline{0.509^{*}}$ \\
Both-Inst & $0.609$ & $0.607$ & \cellcolor{phigh}$0.586^{*}$ & \cellcolor{nmiddle}$0.553^{\bigtriangledown}$ & \cellcolor{nlow}$0.426$ \\
Recency-Focused & \cellcolor{nslow}$0.601$ & \cellcolor{plow}$0.627$ & \cellcolor{plow}$0.565$ & \cellcolor{nlow}$0.562$ & \cellcolor{phigh}$0.481$ \\
RolePlay-User & \cellcolor{nshigh}$0.540^{\bigtriangledown}$ & $0.606$ & \cellcolor{phigh}$0.591^{*}$ & \cellcolor{nshigh}$0.529^{\bigtriangledown}$ & \cellcolor{nlow}$0.428$ \\
RolePlay-Expert & \cellcolor{nslow}$0.592$ & \cellcolor{nslow}$0.597$ & \cellcolor{plow}$0.566$ & \cellcolor{nslow}$0.581$ & \cellcolor{phigh}$0.478$ \\
\bottomrule
\end{tabular}
}
\end{table*}

\begin{table*}[htbp]
\caption{Movies}
\label{table:movie}
\centering
\scalebox{1}{
\begin{tabular}{llllll}
\toprule
 & gpt-4.1-mini & llama3.3-70b & gpt-4o-mini & phi4 & amazon-nova-lite \\
\midrule
Baseline & $0.516$ & $0.506$ & $0.502$ & $0.462$ & $0.401$ \\
Emotion & \cellcolor{nlow}$0.491^{\bigtriangledown}$ & \cellcolor{pslow}$0.514$ & \cellcolor{plow}$0.527$ & \cellcolor{nlow}$0.446$ & \cellcolor{plow}$0.418$ \\
Re-Reading & \cellcolor{nmiddle}$0.487$ & \cellcolor{plow}$0.526$ & \cellcolor{nlow}$0.483$ & \cellcolor{nslow}$0.457$ & \cellcolor{nmiddle}$0.381$ \\
Both-Inst & $0.513$ & $0.502$ & \cellcolor{nlow}$0.484$ & \cellcolor{nslow}$0.455$ & \cellcolor{pslow}$0.413$ \\
Recency-Focused & $0.519$ & \cellcolor{plow}$0.523$ & $0.497$ & \cellcolor{nslow}$0.455$ & \cellcolor{plow}$0.415$ \\
RolePlay-User & \cellcolor{nmiddle}$0.484$ & \cellcolor{pslow}$0.512$ & \cellcolor{nslow}$0.492$ & \cellcolor{nhigh}$0.423^{\bigtriangledown}$ & \cellcolor{phigh}$0.431$ \\
RolePlay-Expert & \cellcolor{nslow}$0.507$ & \cellcolor{pslow}$0.516$ & $0.503$ & $0.461$ & \cellcolor{phigh}$0.440^{*}$ \\
\bottomrule
\bottomrule
\end{tabular}
}
\end{table*}

\begin{table*}[htbp]
\caption{Groceries}
\label{table:groceries}
\centering
\scalebox{1}{
\begin{tabular}{llllll}
\toprule
 & gpt-4.1-mini & llama3.3-70b & gpt-4o-mini & phi4 & amazon-nova-lite \\
\midrule
Baseline & $0.443$ & $0.478$ & $0.440$ & $0.464$ & $0.414$ \\
Emotion & \cellcolor{nslow}$0.438$ & \cellcolor{pslow}$0.484$ & \cellcolor{pslow}$0.445$ & \cellcolor{plow}$0.483$ & \cellcolor{nhigh}$0.382$ \\
Re-Reading & \cellcolor{nmiddle}$0.420$ & \cellcolor{pshigh}$\underline{0.554^{*}}$ & \cellcolor{phigh}$0.471$ & \cellcolor{plow}$0.478$ & \cellcolor{pslow}$\underline{0.421}$ \\
Both-Inst & $0.443$ & \cellcolor{plow}$0.495$ & $0.437$ & $0.462$ & \cellcolor{nhigh}$0.379$ \\
Recency-Focused & \cellcolor{pmiddle}$0.466$ & \cellcolor{plow}$0.495$ & \cellcolor{pmiddle}$0.466$ & \cellcolor{nslow}$0.457$ & $0.412$ \\
RolePlay-User & \cellcolor{nshigh}$0.372^{\bigtriangledown}$ & $0.482$ & \cellcolor{nshigh}$0.358^{\bigtriangledown}$ & \cellcolor{nshigh}$0.395^{\bigtriangledown}$ & \cellcolor{nshigh}$0.331^{\bigtriangledown}$ \\
RolePlay-Expert & $0.441$ & \cellcolor{pslow}$0.490$ & $0.441$ & \cellcolor{pslow}$0.469$ & \cellcolor{nmiddle}$0.387$ \\
\bottomrule
\end{tabular}
}
\end{table*}

\begin{table*}[htbp]
\caption{Clothes}
\label{table:clothes}
\centering
\scalebox{1}{
\begin{tabular}{llllll}
\toprule
 & gpt-4.1-mini & llama3.3-70b & gpt-4o-mini & phi4 & amazon-nova-lite \\
\midrule
Baseline & $0.408$ & $0.408$ & $0.393$ & $\underline{0.471}$ & $0.367$ \\
Emotion & $0.411$ & $0.406$ & \cellcolor{pslow}$0.400$ & \cellcolor{nlow}$0.451$ & \cellcolor{plow}$0.385$ \\
Re-Reading & \cellcolor{nslow}$0.400$ & \cellcolor{phigh}$0.441$ & \cellcolor{nlow}$0.377$ & $0.469$ & \cellcolor{pslow}$0.378$ \\
Both-Inst & \cellcolor{nslow}$0.398$ & $0.404$ & $0.397$ & \cellcolor{nhigh}$0.435^{\bigtriangledown}$ & \cellcolor{pshigh}$0.412^{*}$ \\
Recency-Focused & \cellcolor{pslow}$0.420$ & \cellcolor{pshigh}$\underline{0.473^{*}}$ & \cellcolor{pslow}$0.398$ & \cellcolor{nhigh}$0.429$ & \cellcolor{pshigh}$\underline{0.433^{*}}$ \\
RolePlay-User & \cellcolor{nshigh}$0.325^{\bigtriangledown}$ & \cellcolor{phigh}$0.446^{*}$ & \cellcolor{nshigh}$0.344^{\bigtriangledown}$ & \cellcolor{nshigh}$0.399^{\bigtriangledown}$ & $0.369$ \\
RolePlay-Expert & \cellcolor{nmiddle}$0.380^{\bigtriangledown}$ & \cellcolor{nslow}$0.400$ & \cellcolor{nslow}$0.383$ & \cellcolor{nlow}$0.451$ & \cellcolor{pmiddle}$0.392$ \\
\bottomrule
\end{tabular}
}
\end{table*}

\begin{table*}[htbp]
\caption{Books}
\label{table:books}
\centering
\scalebox{1}{
\begin{tabular}{llllll}
\toprule
 & gpt-4.1-mini & llama3.3-70b & gpt-4o-mini & phi4 & amazon-nova-lite \\
\midrule
Baseline & $0.585$ & $0.574$ & $0.518$ & $0.595$ & $0.470$ \\
Emotion & $0.582$ & $0.575$ & \cellcolor{plow}$0.543$ & \cellcolor{nslow}$0.580$ & \cellcolor{nslow}$0.461$ \\
Re-Reading & \cellcolor{nmiddle}$0.553^{\bigtriangledown}$ & $0.577$ & \cellcolor{nshigh}$0.436^{\bigtriangledown}$ & \cellcolor{nshigh}$0.533^{\bigtriangledown}$ & \cellcolor{pmiddle}$0.494$ \\
Both-Inst & \cellcolor{nlow}$0.566$ & \cellcolor{nslow}$0.561$ & $0.520$ & \cellcolor{nslow}$0.580$ & \cellcolor{pslow}$0.475$ \\
Recency-Focused & \cellcolor{pslow}$0.591$ & \cellcolor{plow}$0.601$ & \cellcolor{pslow}$0.532$ & $\underline{0.598}$ & \cellcolor{nlow}$0.450$ \\
RolePlay-User & \cellcolor{nhigh}$0.533^{\bigtriangledown}$ & \cellcolor{nslow}$0.560$ & \cellcolor{pmiddle}$0.545$ & \cellcolor{nlow}$0.573$ & \cellcolor{nhigh}$0.426$ \\
RolePlay-Expert & $0.587$ & \cellcolor{nslow}$0.561^{\bigtriangledown}$ & \cellcolor{nslow}$0.513$ & $0.592$ & \cellcolor{nmiddle}$0.442$ \\
\bottomrule
\end{tabular}

}
\end{table*}

\begin{table*}[htbp]
\caption{Yelp}
\label{table:yelp}
\centering
\scalebox{1}{
\begin{tabular}{llllll}
\toprule
 & gpt-4.1-mini & llama3.3-70b & gpt-4o-mini & phi4 & amazon-nova-lite \\
\midrule
Baseline & $0.512$ & $0.475$ & $0.501$ & $0.331$ & $\underline{0.447}$ \\
Emotion & \cellcolor{nslow}$0.503$ & \cellcolor{pslow}$0.484$ & \cellcolor{nlow}$0.479$ & \cellcolor{pmiddle}$0.352$ & \cellcolor{nhigh}$0.404^{\bigtriangledown}$ \\
Re-Reading & \cellcolor{nshigh}$0.421^{\bigtriangledown}$ & \cellcolor{pmiddle}$0.507^{*}$ & \cellcolor{nshigh}$0.439^{\bigtriangledown}$ & \cellcolor{nhigh}$0.306$ & \cellcolor{nshigh}$0.389^{\bigtriangledown}$ \\
Both-Inst & \cellcolor{pmiddle}$0.544^{*}$ & \cellcolor{pmiddle}$0.507^{*}$ & \cellcolor{nlow}$0.485$ & \cellcolor{phigh}$0.360$ & \cellcolor{nlow}$0.433$ \\
Recency-Focused & \cellcolor{pslow}$0.526$ & \cellcolor{nslow}$0.468$ & \cellcolor{nlow}$0.480$ & \cellcolor{pmiddle}$0.352$ & \cellcolor{nshigh}$0.402^{\bigtriangledown}$ \\
RolePlay-User & \cellcolor{nslow}$0.500$ & \cellcolor{pslow}$0.481$ & \cellcolor{nslow}$0.493$ & \cellcolor{nlow}$0.320$ & \cellcolor{nlow}$0.425$ \\
RolePlay-Expert & $0.515$ & \cellcolor{plow}$0.490$ & $0.505$ & \cellcolor{nslow}$0.322$ & \cellcolor{nshigh}$0.397^{\bigtriangledown}$ \\
\bottomrule
\end{tabular}
}
\end{table*}

\begin{table*}[htbp]
\caption{News}
\label{table:news}
\centering
\scalebox{1}{
\begin{tabular}{llllll}
\toprule
 & gpt-4.1-mini & llama3.3-70b & gpt-4o-mini & phi4 & amazon-nova-lite \\
\midrule
Baseline & $0.262$ & $0.157$ & $0.200$ & $0.145$ & $0.267$ \\
Emotion & \cellcolor{plow}$0.271$ & \cellcolor{pshigh}$0.177$ & \cellcolor{pslow}$0.204$ & \cellcolor{nhigh}$0.131$ & $0.265$ \\
Re-Reading & \cellcolor{plow}$0.275$ & \cellcolor{pshigh}$0.233^{*}$ & \cellcolor{phigh}$0.219$ & \cellcolor{pshigh}$0.244^{*}$ & \cellcolor{pshigh}$0.316^{*}$ \\
Both-Inst & \cellcolor{nshigh}$0.226^{\bigtriangledown}$ & \cellcolor{phigh}$0.169$ & \cellcolor{nslow}$0.194$ & \cellcolor{nshigh}$0.083^{\bigtriangledown}$ & \cellcolor{pshigh}$0.383^{*}$ \\
Recency-Focused & \cellcolor{nmiddle}$0.245$ & \cellcolor{pshigh}$0.248^{*}$ & \cellcolor{pshigh}$0.230$ & \cellcolor{nhigh}$0.131$ & \cellcolor{pslow}$0.272$ \\
RolePlay-User & \cellcolor{nshigh}$0.195^{\bigtriangledown}$ & \cellcolor{phigh}$0.169$ & \cellcolor{nshigh}$0.116^{\bigtriangledown}$ & \cellcolor{nshigh}$0.042^{\bigtriangledown}$ & \cellcolor{pshigh}$0.439^{*}$ \\
RolePlay-Expert & $0.260$ & \cellcolor{nslow}$0.154$ & \cellcolor{nlow}$0.191$ & \cellcolor{nhigh}$0.132$ & \cellcolor{nlow}$0.254$ \\
\bottomrule
\end{tabular}
}
\end{table*}

\begin{table*}[htbp]
\caption{Food}
\label{table:food}
\centering
\scalebox{1}{
\begin{tabular}{llllll}
\toprule
 & gpt-4.1-mini & llama3.3-70b & gpt-4o-mini & phi4 & amazon-nova-lite \\
\midrule
Baseline & $0.328$ & $0.312$ & $0.364$ & $0.278$ & $0.305$ \\
Emotion & $0.327$ & \cellcolor{nmiddle}$0.290$ & $0.360$ & \cellcolor{pshigh}$\underline{0.319^{*}}$ & \cellcolor{nshigh}$0.266$ \\
Re-Reading & \cellcolor{nhigh}$0.304$ & \cellcolor{nslow}$0.308$ & \cellcolor{nhigh}$0.335$ & \cellcolor{nshigh}$0.244$ & \cellcolor{nshigh}$0.258$ \\
Both-Inst & \cellcolor{nlow}$0.316$ & \cellcolor{nshigh}$0.248^{\bigtriangledown}$ & \cellcolor{nmiddle}$0.341$ & \cellcolor{pmiddle}$0.293$ & \cellcolor{nshigh}$0.249^{\bigtriangledown}$ \\
Recency-Focused & \cellcolor{pshigh}$\underline{0.361}$ & \cellcolor{plow}$0.324$ & \cellcolor{plow}$0.375$ & \cellcolor{pmiddle}$0.292$ & \cellcolor{nmiddle}$0.289$ \\
RolePlay-User & \cellcolor{nlow}$0.317$ & \cellcolor{pslow}$0.315$ & \cellcolor{nlow}$0.348$ & \cellcolor{pshigh}$0.316$ & \cellcolor{nhigh}$0.280$ \\
RolePlay-Expert & \cellcolor{nhigh}$0.305$ & $0.313$ & \cellcolor{nmiddle}$0.344$ & \cellcolor{pshigh}$0.310^{*}$ & \cellcolor{nshigh}$0.273$ \\
\bottomrule
\end{tabular}
}
\end{table*}

\begin{table*}[htbp]
\caption{Light}
\label{table:light}
\centering
\scalebox{1}{
\begin{tabular}{llllll}
\toprule
 & gpt-4.1-mini & llama3.3-70b & gpt-4o-mini & phi4 & amazon-nova-lite \\
\midrule
Baseline & $0.460$ & $0.440$ & $0.444$ & $0.423$ & $0.360$ \\
Emotion & $0.461$ & $0.439$ & \cellcolor{pslow}$0.450$ & \cellcolor{pslow}$0.428$ & \cellcolor{nlow}$0.347$ \\
Re-Reading & \cellcolor{nlow}$0.444$ & \cellcolor{phigh}$0.471^{*}$ & \cellcolor{nlow}$0.431$ & \cellcolor{pslow}$0.432$ & \cellcolor{pmiddle}$0.379$ \\
Both-Inst & $0.459$ & $0.437$ & \cellcolor{nslow}$0.432^{\bigtriangledown}$ & \cellcolor{nslow}$0.417$ & \cellcolor{pslow}$0.364$ \\
Recency-Focused & \cellcolor{plow}$0.478$ & \cellcolor{pmiddle}$0.467^{*}$ & \cellcolor{pslow}$0.449$ & $0.424$ & \cellcolor{nslow}$0.357$ \\
RolePlay-User & \cellcolor{nhigh}$0.420^{\bigtriangledown}$ & \cellcolor{plow}$0.461^{*}$ & \cellcolor{nlow}$0.426^{\bigtriangledown}$ & \cellcolor{nhigh}$0.391^{\bigtriangledown}$ & \cellcolor{plow}$0.372$ \\
RolePlay-Expert & \cellcolor{nslow}$0.450$ & $0.439$ & $0.443$ & $0.423$ & $0.358$ \\
\bottomrule
\end{tabular}
}
\end{table*}

\begin{table*}[htbp]
\caption{Heavy}
\label{table:heavy}
\centering
\scalebox{1}{
\begin{tabular}{llllll}
\toprule
 & gpt-4.1-mini & llama3.3-70b & gpt-4o-mini & phi4 & amazon-nova-lite \\
\midrule
Baseline & $0.456$ & $0.439$ & $0.421$ & $0.411$ & $0.418$ \\
Emotion & \cellcolor{nslow}$0.447$ & \cellcolor{pslow}$0.445$ & \cellcolor{pslow}$0.430$ & \cellcolor{nslow}$0.406$ & \cellcolor{nlow}$0.404$ \\
Re-Reading & \cellcolor{nhigh}$0.412^{\bigtriangledown}$ & \cellcolor{phigh}$0.474^{*}$ & \cellcolor{nshigh}$0.374^{\bigtriangledown}$ & \cellcolor{nmiddle}$0.386^{\bigtriangledown}$ & \cellcolor{nslow}$0.408$ \\
Both-Inst & \cellcolor{nslow}$0.444$ & $0.436$ & \cellcolor{pslow}$0.429$ & \cellcolor{nmiddle}$0.388^{\bigtriangledown}$ & \cellcolor{pslow}$0.429$ \\
Recency-Focused & $0.455$ & \cellcolor{phigh}$0.473^{*}$ & \cellcolor{plow}$0.437$ & \cellcolor{nlow}$0.395$ & \cellcolor{plow}$\underline{0.432}$ \\
RolePlay-User & \cellcolor{nshigh}$0.397^{\bigtriangledown}$ & \cellcolor{nslow}$0.432$ & \cellcolor{nmiddle}$0.396^{\bigtriangledown}$ & \cellcolor{nshigh}$0.358^{\bigtriangledown}$ & \cellcolor{nslow}$0.411$ \\
RolePlay-Expert & \cellcolor{nslow}$0.446$ & $0.441$ & $0.418$ & \cellcolor{nslow}$0.406$ & \cellcolor{nslow}$0.408$ \\
\bottomrule
\end{tabular}
}
\end{table*}

\begin{table*}[htbp]
\caption{Hit@3 for Table~7 in the main paper.}
\label{table:hit3}
\centering
\scalebox{0.75}{
\begin{tabular}{llllllllllllllll}
\toprule
 & Music & Movie & Groceries & Clothes & Book & Yelp & News & Food & Light & Heavy & 4.1-mini & llama3.3 & 4o-mini & phi4 & nova-lite \\
\midrule
Baseline & $0.720$ & $0.595$ & $0.525$ & $0.520$ & $0.675$ & $0.680$ & $0.380$ & $0.445$ & $0.568$ & $0.568$ & $0.568$ & $0.543$ & $0.537$ & $0.514$ & $0.497$ \\
Emotion & $0.715$ & \cellcolor{nmiddle}$0.565^{\bigtriangledown}$ & \cellcolor{pslow}$0.535$ & \cellcolor{nslow}$0.505$ & $0.675$ & \cellcolor{nslow}$0.665$ & \cellcolor{pslow}$0.390$ & \cellcolor{nslow}$0.440$ & $0.569$ & \cellcolor{nslow}$0.554$ & \cellcolor{nslow}$0.561$ & \cellcolor{pslow}$0.551$ & \cellcolor{pslow}$0.543$ & \cellcolor{pslow}$0.519$ & \cellcolor{nlow}$0.480^{\bigtriangledown}$ \\
Re-Reading & \cellcolor{nlow}$0.690$ & \cellcolor{pslow}$0.605$ & $0.520$ & \cellcolor{nlow}$0.500$ & \cellcolor{nmiddle}$0.630^{\bigtriangledown}$ & \cellcolor{nshigh}$0.545^{\bigtriangledown}$ & \cellcolor{pslow}$0.390$ & \cellcolor{nshigh}$0.400$ & \cellcolor{nslow}$0.555$ & \cellcolor{nhigh}$0.515^{\bigtriangledown}$ & \cellcolor{nmiddle}$0.535^{\bigtriangledown}$ & \cellcolor{phigh}$0.586^{*}$ & \cellcolor{nhigh}$0.493^{\bigtriangledown}$ & $0.509$ & \cellcolor{pslow}$\underline{0.511}$ \\
Both-Inst & $0.715$ & $0.595$ & \cellcolor{plow}$0.550$ & \cellcolor{nslow}$0.505$ & \cellcolor{nlow}$0.645$ & \cellcolor{plow}$\underline{0.710}$ & \cellcolor{nshigh}$0.340^{\bigtriangledown}$ & \cellcolor{nlow}$0.425$ & $0.570$ & \cellcolor{nslow}$0.551$ & \cellcolor{nslow}$0.561$ & $0.540$ & \cellcolor{nslow}$0.524$ & \cellcolor{nlow}$0.497$ & \cellcolor{pslow}$0.505$ \\
Recency-Focused & \cellcolor{nlow}$0.695$ & $0.590$ & \cellcolor{pmiddle}$0.555$ & $0.525$ & \cellcolor{nslow}$0.665$ & $0.680$ & \cellcolor{nmiddle}$0.360$ & \cellcolor{pmiddle}$\underline{0.470}$ & \cellcolor{pslow}$0.580$ & \cellcolor{nslow}$0.555$ & $0.568$ & \cellcolor{plow}$0.570^{*}$ & $0.539$ & $0.511$ & \cellcolor{pslow}$0.504$ \\
\bottomrule
\end{tabular}
}
\end{table*}

\begin{table*}[htbp]
\caption{Hit@3 for Table~13 in the main paper.}
\label{table:hit3_high}
\centering
\scalebox{1}{
\begin{tabular}{lllll}
\toprule
 & Baseline & Rephrase & Step-Back & ReAct \\
\midrule
llama3.3-70b & $0.537$ & \cellcolor{pshigh}$0.650^{*}$ & \cellcolor{pshigh}$\underline{0.656^{*}}$ & \cellcolor{pshigh}$0.644^{*}$ \\
gpt-4.1-mini & $0.581$ & \cellcolor{pshigh}$\underline{0.662^{*}}$ & \cellcolor{pmiddle}$0.619$ & \cellcolor{pshigh}$0.650^{*}$ \\
claude-3.5-haiku & $0.606$ & \cellcolor{pslow}$0.613$ & \cellcolor{nslow}$0.600$ & \cellcolor{plow}$\underline{0.625}$ \\
gpt-4.1 & $0.619$ & \cellcolor{nslow}$0.606$ & \cellcolor{nslow}$0.606$ & \cellcolor{plow}$\underline{0.644}$ \\
claude-3.7-sonnet & $0.738$ & \cellcolor{nslow}$0.719$ & $\underline{0.744}$ & \cellcolor{nlow}$0.706$ \\
o3-mini & $\underline{0.675}$ & \cellcolor{nmiddle}$0.631$ & \cellcolor{nmiddle}$0.637$ &  \\
o4-mini & $\underline{0.656}$ &  &  &  \\
claude-3.7-sonnet (T) & $\underline{0.637}$ &  &  &  \\
o3 & $\underline{0.725}$ &  &  &  \\
\bottomrule
\end{tabular}

}
\end{table*}

\clearpage

\begin{table*}[htbp]
\caption{Percentages of outputs with containing five or fewer ranked item (left) and those containing nine or fewer ranked items (right).}
\label{table:partial_ranking_1_5}
\centering
\scalebox{1}{
\begin{tabular}{llllll}
\toprule
 & gpt-4.1-mini & llama3.3-70b & gpt-4o-mini & phi4 & amazon-nova-lite \\
\midrule
Baseline & $0 / 0.1$ & $0 / 0.1$ & $4.9 / 5.9$ & $\textbf{12.9} / \textbf{13.5}$ & $0.6 / 0.8$ \\
Emotion & $0 / 0.1$ & $0 / 0.1$ & $4.1 / 4.9$ & $\textbf{12.8} / \textbf{13.0}$ & $0.2 / 0.4$ \\
Re-Reading & $0 / 0.1$ & $0 / 0.1$ & $\textbf{15.6} / \textbf{18.0}$ & $\textbf{27.8} / \textbf{28.3}$ & $0 / 0.1$ \\
Both-Inst & $0 / 0.1$ & $0 / 0.1$ & $2.8 / 3.6$ & $\textbf{13.4} / \textbf{13.5}$ & $0.6 / 0.8$ \\
Recency-Focused & $0 / 0.1$ & $0 / 0.1$ & $6.2 / 6.9$ & $\textbf{12.9} / \textbf{12.9}$ & $0.5 / 0.6$ \\
RolePlay-User & $0 / 0.1$ & $0 / 0.1$ & $0.4 / 0.8$ & $\textbf{12.6} / \textbf{12.9}$ & $0.1 / 0.2$ \\
RolePlay-Expert & $0 / 0.1$ & $0 / 0.1$ & $4.5 / 5.4$ & $\textbf{12.5} / \textbf{12.9}$ & $1.9 / 2.0$ \\
\bottomrule
\end{tabular}
}
\end{table*}

\begin{table*}[htbp]
\caption{Token Number Statistics (Baseline)}
\label{table:token_stat_baseline}
\centering
\scalebox{1}{
\begin{tabular}{l|llll|llll}
\toprule
 & \multicolumn{4}{c|}{Light} & \multicolumn{4}{c}{Heavy} \\
 & max & min & mean & std & max & min & mean & std \\
\midrule
Yelp & 9662 & 2959 & 6218.2 & 1199.7 & 48660 & 13363 & 24749.6 & 4674.3 \\
News & 1966 & 1227 & 1537.3 & 151.3 & 5516 & 2932 & 3890.8 & 495.5 \\
Food & 6218 & 3991 & 4859.1 & 353.2 & 18552 & 12373 & 15463.6 & 1117.0 \\
Movie & 3573 & 1121 & 2146.0 & 487.3 & 13282 & 3173 & 6793.2 & 1964.9 \\
Music & 3884 & 1038 & 1943.1 & 465.5 & 12315 & 2767 & 6579.7 & 2180.9 \\
Groceries & 4427 & 1668 & 2921.1 & 533.8 & 15994 & 5453 & 9414.0 & 2211.5 \\
Clothes & 4050 & 1340 & 2571.5 & 486.4 & 14379 & 3905 & 7168.7 & 1794.8 \\
Book & 4316 & 1004 & 1896.2 & 527.5 & 12014 & 3007 & 6477.8 & 1764.1 \\
\bottomrule
\end{tabular}
}
\end{table*}

\begin{table*}[htbp]
\caption{Token Number Statistics (Summarize-Item)}
\label{table:token_stat_itemsummary}
\centering
\scalebox{1}{
\begin{tabular}{l|llll|llll}
\toprule
 & \multicolumn{4}{c|}{Light} & \multicolumn{4}{c}{Heavy} \\
 & max & min & mean & std & max & min & mean & std \\
\midrule
Yelp & 7735 & 3464 & 4685.1 & 703.0 & 41732 & 12718 & 17863.0 & 3784.7 \\
News & 2670 & 1894 & 2285.6 & 176.9 & 7277 & 4776 & 5974.3 & 650.0 \\
Food & 5461 & 3716 & 4198.0 & 271.3 & 16199 & 10683 & 13388.6 & 1029.1 \\
Movie & 4807 & 2299 & 3050.8 & 435.3 & 16693 & 6427 & 9431.4 & 2106.2 \\
Music & 4936 & 2317 & 3048.5 & 452.6 & 16820 & 6124 & 9952.4 & 2374.8 \\
Groceries & 4160 & 2426 & 3064.5 & 398.0 & 16654 & 6605 & 9706.0 & 2059.7 \\
Clothes & 4328 & 2125 & 2751.8 & 384.5 & 14796 & 5884 & 8092.4 & 1588.4 \\
Book & 5287 & 2369 & 3104.1 & 517.2 & 16754 & 6367 & 9881.5 & 1995.0 \\
\bottomrule
\end{tabular}
}
\end{table*}

\begin{figure*}[htbp]
\centering
\begin{lstlisting}[frame=single]
(*@{$t_{a1}$@*) = ### Observation:
The user has a history of reviewing ballet and dance-related workout DVDs. They express a strong preference for challenging workouts that provide effective results, particularly in toning and defining muscles. The user has rated several products, with the highest rating being 5.0 for a challenging ballet workout, while others received lower ratings due to being too easy or not effective enough.

### Thought:
The user seems to prefer products that are intense and provide a good workout, particularly in ballet and dance fitness. They have explicitly mentioned the need for challenging workouts that yield visible results. Therefore, any candidate products that align with these preferences, especially in the fitness category, would likely be more appealing to the user.

### Action:
I will examine the candidate products to identify any that align with the user's preferences for challenging workouts, particularly in the exercise and fitness category.

1. **Emmet Otter's Jug-Band Christmas** - Not related to fitness.
(*@\textcolor{blue}{(omitted)}@*)
10. **Royal Paintbox** - Not related to fitness.

### Observation:
Among the candidate products, only **Tracy Anderson: Precision Toning** is related to fitness. However, based on the user's previous experiences with Tracy Anderson's workouts, they found them too easy. The other products do not align with the user's interests in ballet or challenging fitness routines.

### Thought:
Since there is only one candidate product related to fitness, and it does not align with the user's preference for challenging workouts, I must conclude that there are no suitable recommendations from the provided candidates.

### Answer:
Since none of the candidate products align with the user's preferences for challenging ballet or dance workouts, the final answer is: **[]** (no suitable recommendations).
\end{lstlisting}
\caption{Incorrect output generated by the ReAct prompt}
\label{list:fail_ReAct}
\end{figure*}

\clearpage

\begin{table*}[htbp]
\caption{Inference Cost for 1,600 users in Section 4}
\label{table:cost_section4}
\centering
\scalebox{1}{
\begin{tabular}{llllll}
\toprule
 & gpt-4.1-mini & llama3.3-70b & gpt-4o-mini & phi4 & amazon-nova-lite \\
\midrule
Baseline & 4.25 & 7.73 & 1.61 & 1.71 & 0.70 \\
Emotion & 4.27 & 7.76 & 1.61 & 1.71 & 0.70 \\
Re-Reading & 8.44 & 15.40 & 3.19 & 4.93 & 1.40 \\
Both-Inst & 4.25 & 7.73 & 1.61 & 1.67 & 0.70 \\
Recency-Focused & 4.38 & 7.97 & 1.68 & 1.79 & 0.72 \\
RolePlay-User & 4.23 & 7.71 & 1.59 & 1.64 & 0.69 \\
RolePlay-Expert & 4.25 & 7.73 & 1.61 & 1.70 & 0.70 \\
RolePlay-Frederick & 4.26 & 7.73 & 1.61 & 1.71 & 0.70 \\
Summarize-Item & 4.48 & 8.11 & 1.69 & 1.20 & 0.70 \\
Step-Back & 10.66 & 16.82 & 3.77 & 3.13 & 2.87 \\
ReAct & 11.76 & 17.70 & 4.39 & 3.05 & 2.47 \\
Rephrase & 11.59 & 17.35 & 3.87 & 2.98 & 1.99 \\
Echo & 8.40 & 15.34 & 3.19 & 4.83 & 1.39 \\
Summarize-User & 10.55 & 16.74 & 3.92 & 2.81 & 1.75 \\
Generate-Item & 8.58 & 15.62 & 3.41 & 2.67 & 1.41 \\
Reuse-Item & 4.67 & 8.26 & 1.71 & 1.88 & 0.75 \\
Explain & 10.09 & 16.60 & 3.81 & 2.81 & 1.65 \\
Mock & 4.33 & 7.88 & 1.63 & 1.74 & 0.71 \\
Step-by-Step & 8.60 & 15.67 & 3.31 & 3.14 & 1.54 \\
Deep-Breath & 8.61 & 15.68 & 3.28 & 3.41 & 1.47 \\
Plan-Solve & 8.63 & 15.73 & 3.34 & 3.71 & 2.19 \\
sum & 149.28 & 257.27 & 55.83 & 54.21 & 27.21 \\
\bottomrule
\end{tabular}
}
\end{table*}

\begin{table*}[htbp]
\caption{Inference Cost for 160 users in Section 5.1}
\label{table:cost_section51}
\centering
\scalebox{1}{
\begin{tabular}{llllll}
\toprule
 & gpt-4.1-mini & llama3.3-70b & gpt-4o-mini & phi4 & amazon-nova-lite \\
\midrule
Baseline & 0.43 & 0.78 & 0.33 & 0.17 & 0.14 \\
Rephrase & 1.15 & 1.74 & 0.38 & 0.28 & 0.20 \\
Step-Back & 1.08 & 1.69 & 0.38 & 0.32 & 0.30 \\
ReAct & 1.18 & 1.78 & 0.44 & 0.30 & 0.26 \\
Rephrase$\rightarrow$ReAct & 1.87 & 2.88 & 0.75 & 0.46 & 0.38 \\
Rephrase$\rightarrow$Step-Back & 1.72 & 2.82 & 0.66 & 0.44 & 0.36 \\
SelfRefine (Rephrase) & 1.73 & 2.77 & 0.62 & 0.41 & 0.32 \\
SelfRefine (Step-Back) & 1.73 & 2.68 & 0.60 & 0.48 & 0.38 \\
SelfRefine (ReAct) & 1.83 & 2.79 & 0.69 & 0.51 & 0.33 \\
SelfConsistency (Rephrase) & 1.75 & 2.74 & 0.69 & 0.42 & 0.33 \\
SelfConsistency (Step-Back) & 1.76 & 2.63 & 0.60 & 0.46 & 0.26 \\
SelfConsistency (ReAct) & 1.95 & 2.77 & 0.70 & 0.43 & 0.30 \\
sum & 18.17 & 28.06 & 6.84 & 4.69 & 3.54 \\
\bottomrule
\end{tabular}

}
\end{table*}

\begin{table*}[htbp]
\caption{Inference Cost for 160 users in Section 5.2}
\label{table:cost_section52}
\centering
\scalebox{1}{
\begin{tabular}{llllll}
\toprule
 & Baseline & ReAct & Rephrase & Step-Back & sum \\
\midrule
phi4 & 0.17 & 0.30 & 0.28 & 0.32 & 1.07 \\
amazon-nova-lite & 0.14 & 0.26 & 0.20 & 0.30 & 0.90 \\
gpt-4o-mini & 0.33 & 0.44 & 0.38 & 0.38 & 1.53 \\
llama3.3-70b & 0.78 & 1.78 & 1.74 & 1.69 & 5.98 \\
gpt-4.1-mini & 0.43 & 1.18 & 1.15 & 1.08 & 3.83 \\
claude-3.5-haiku & 1.03 & 2.41 & 2.28 & 2.30 & 8.02 \\
gpt-4.1 & 2.13 & 5.76 & 4.89 & 4.87 & 17.65 \\
claude-3.7-sonnet & 3.85 & 10.69 & 9.23 & 8.71 & 32.48 \\
o3-mini & 2.79 & - & 5.00 & 4.82 & 12.62 \\
o4-mini & 2.09 & - & - & - & 2.09 \\
claude-3.7-sonnet (T) & 7.72 & - & - & - & 7.72 \\
o3 & 11.36 & - & - & - & 11.36 \\
\bottomrule
\end{tabular}
}
\end{table*}

\end{document}